\documentclass[preprint,12pt]{elsarticle}
\usepackage{amssymb}
\usepackage{lineno}

\usepackage{times}

\biboptions{round}
\biboptions{semicolon}
\biboptions{authoryear}

\usepackage{booktabs}
\usepackage{amsmath}
\usepackage{hyperref}
\hypersetup{
    colorlinks=true,
    linkcolor=blue,
    filecolor=magenta,      
    urlcolor=cyan,
    pdftitle={Sharelatex Example},
    pdfpagemode=FullScreen,
    }

\usepackage{caption}
\usepackage{xcolor}

\usepackage[a4paper, portrait, margin=1.25in]{geometry}
\usepackage{multirow}

\usepackage{threeparttable}

\usepackage{hyphenat} 

\usepackage[acronym]{glossaries}
\makeglossaries
\newacronym{dcm}{DCM}{Discrete Choice Model}
\newacronym{rum}{RUM}{Random Utility Model}
\newacronym{msl}{MSL}{Maximum Simulated Likelihood}
\newacronym{abm}{ABM}{Activity-Based Model}
\newacronym{mnl}{MNL}{Multinomial Logit}
\newacronym{lcm}{LCM}{Latent Class Model}
\newacronym{gl}{GL}{General Loss}
\newacronym{mcmc}{MCMC}{Markov Chain Monte Carlo}
\newacronym{ml}{ML}{Machine Learning}
\newacronym{ai}{AI}{Artificial Intelligence}
\newacronym{ltds}{LTDS}{London Travel Demand Survey}
\newacronym{luti}{LUTI}{London Land-Use and Transport Information}
\newacronym{bda}{BDA}{Bayesian Data Assimilation}
\newacronym{esbda}{ESBDA}{Early Stopping Bayesian Data Assimilation}
\newacronym{mmnl}{MMNL}{Mixed Multinomial Logit}
\newacronym{api}{API}{application programming interface}
\newacronym{iid}{iid}{independent and identically distributed}
\newacronym{cel}{CEL}{Cross-Entropy Loss}
\newacronym{hb}{HB}{hierarchical Bayes}
\newacronym{ev1}{EV1}{Extreme-Value 1 type}
\newacronym{ll}{LL}{log-likelihood}
\newacronym{mh}{M-H}{Metropolis–Hastings Algorithm}
\newacronym{vot}{VOT}{Value of Time}
\newacronym{gmpca}{GMPCA}{Geometric Mean Probability of Correct Assignment}
\newacronym{nl}{NL}{Nested Logit}
\newacronym{ol}{OL}{Ordered Logit}

\usepackage{lipsum}
\makeatletter
\def\ps@pprintTitle{%
 \let\@oddhead\@empty
 \let\@evenhead\@empty
 \def\@oddfoot{}%
 \let\@evenfoot\@oddfoot}
\makeatother

\begin{document}

\begin{frontmatter}

\title{An Early Stopping Bayesian Data Assimilation Approach for Mixed-Logit Estimation}

\author{Shanshan Xie\corref{cor1}}
\ead{sx239@cam.ac.uk}
\address{Department of Architecture, 1-5 Scroope Terrace, Cambridge, CB2 1PX, the United Kingdom}
\author{Tim Hillel}
\ead{tim.hillel@epfl.ch}
\address{Transportation and Mobility Laboratory, ENAC IIC TRANSP-OR, GC B2 402 (Bâtiment GC), Station 18, CH-1015 Lausanne, Switzerland}
\author{Ying Jin}
\ead{Ying.Jin@aha.cam.ac.uk}
\address{Department of Architecture, 1-5 Scroope Terrace, Cambridge, CB2 1PX, the United Kingdom}

\begin{abstract}

The mixed-logit model is a flexible tool in transportation choice analysis, 
which provides valuable insights into inter and intra-individual behavioural heterogeneity. 
However, applications of mixed-logit models in practice are limited by the
high computational and data requirements for model estimation.
When estimating mixed-logit models on small samples, the \emph{Bayesian} estimation approach becomes vulnerable to over and under-fitting.
This is problematic for investigating the behaviour of specific population 
sub-groups or market segments, where a modeller may wish to estimate separate 
models for multiple similar contexts, each with low data availability. 
Similar challenges arise when adapting an existing model to a new 
location or time period, e.g., when estimating post-pandemic travel behaviour.

In order to address the data and transferability issues of the mixed-logit 
model, in this paper we propose a new \gls{esbda} simulator 
for estimation of mixed-logit which combines a \emph{Bayesian} statistical approach 
with \gls{ml} model estimation methodologies. 
The \gls{esbda} simulator is intended to improve the transferability of 
mixed-logit models and to enable the estimation of robust choice models with low data availability. 
This approach can therefore provide new insights into people's choice 
behaviour where the traditional estimation of full mixed-logit models was not 
previously possible due to low data availability, and open up new opportunities for investment and 
planning decisions support.  

To assess the performance of the new approach, we benchmark the \gls{esbda} 
estimator against the Direct Application approach and two reference simulators:
(i) a basic \gls{hb} simulator with random starting parameter values; 
(ii) a \gls{bda} simulator without \emph{early stopping}.
Two case-studies are used to investigate the relative performance 
of the simulators in varied contexts.

The experimental results show that the proposed \gls{esbda} approach can 
effectively overcome under and over-fitting and non-convergence issues in 
simulation.
The resulting models from \gls{esbda} clearly outperform those of the 
reference simulators in terms of predictive accuracy. 
Furthermore, we find that models estimated with \gls{esbda} tend to be more 
robust, with significant parameters with signs and values consistent with 
behavioural theory, even when estimated on small samples. 

\end{abstract}


\end{frontmatter}

\printglossary[type=\acronymtype]

\glsresetall
\section{Introduction}
\label{sec: intro}

\glspl{dcm} based on random utility theory are a key research tool in 
behaviour analysis in areas such as transportation, economics, health, and 
many other disciplines where there is a focus on individual choice behaviour. 
Traditional \glspl{dcm}, including the logit and \gls{nl} models, make use of
closed-form \emph{utility specifications} with fixed parameters. 
This allows choice probabilities to be generated analytically without the need 
for simulation.
Despite their computational convenience, the fixed parameters used in logit models 
do not account for the significant inter and intra-individual 
heterogeneity in individual decision-makers' behaviours. 

Mixed-logit models accommodate heterogeneity in \glspl{dcm} by treating model 
parameters as distributed over the population of 
interest \citep[see][]{bhat2000flexible,mcfadden2000mixed,train2003discrete}.
Whilst they extend the behavioural capabilities of logit models, mixed-logit 
models have large data and computational requirements \citep{Greene2003}. 
There are two predominant estimation techniques for mixed-logit models: 
the \gls{msl} method and the \emph{Bayes} approach. 
Both techniques can face issues with convergence and model fitting
errors with low data availability \citep{bhat2001quasi,wang2006convergence}.

These issues can be problematic when investigating the behaviour of specific 
population sub-groups or market segments, where a modeller may want to estimate
separate models for several similar contexts, each with low data availability. 
The same challenges arise when estimating a new model for a poorly sampled 
location or time period, where there may not be sufficient data to estimate
a new model.
In the meantime, empirical models are found to have low \emph{transferability};
i.e. a model estimated in one context cannot accurately predict similar choices for a new context \citep{koppelman1985geographic}.
This means that areas, population segments or new time periods with poor data availability cannot benefit substantially from existing models.
There is therefore the need for an advanced simulation technique for 
mixed-logit estimation to address modelling anomalies caused by data shortage.

This study introduces and evaluates a new estimation approach for improving 
the predictive power and behavioural consistency of mixed-logit models for 
small sample modelling.
The new approach, called \gls{esbda}, is developed to adapt a previously 
established model to a new population subgroup, location, or time period, with 
low data availability. 
This work builds on existing concepts of model transferability 
\citep{Ben-Akiva1987} and the Hierarchical Bayes simulation \citep{Train2006}. 
The configuration of \gls{esbda} aims to address the recurrent issues of 
over-fitting and non-convergence in small sample modelling of mixed-logit 
models.
The primary intended contribution of \gls{esbda} is the improved 
transferability of mixed-logit models. 
This enables an inexpensive and practical way to estimate models for a new 
context.

In addition to tackling locational or demographic heterogeneity, the increased 
model transferability also enables existing models to be better adapted to 
future scenarios and\slash or demographic changes. 
This is increasingly relevant, given the current rate of technological 
innovation, urban demographic changes and the pandemic shock. 
One potential application is the estimation of post-pandemic travel choice behaviour in a fast, inexpensive manner with relatively low data requirements.

\section{Literature Review}
\label{sec: literature}

\subsection{Heterogeneity in Discrete Choice Models}
\label{heterogeneity}

There are many potential sources of heterogeneity in choice behaviour.
Those identified in the literature include: 
difference in behavioural processes, knowledge (e.g., awareness, information) 
and points of view (e.g., perceptions, attitudes and cultural values) between 
individuals \citep{wood2000attitude,ajzen2001nature,sparks2013online}; 
different circumstances between individual choices 
\citep{engel1968consumer,engel1982consumer,engel1995consumer,assael1995consumer,paulssen2014values}; 
and difference in the degree of familiarity, complexity, and perceived risks 
of the choice-set \citep{ajzen1987attitudes,garling1998theoretical,beinamin2013behavioral}.
Whilst the simplicity of logit models enables straightforward model 
estimation, the fixed parameters used in the utility specifications do not 
account for this heterogeneity. 

There have been many proposed variants of the standard logit model to 
accommodate heterogeneity in choice behaviour. 
The predominant approach explored in the literature is to introduce flexibilty 
in the model formulation by allowing individual parameters to be distributed 
across the modelled population. 
The parameter distributions can be discrete, as in the \gls{lcm} 
\citep{bhat1997endogenous,Greene2003}; 
or continuous, as in the Mixed-Logit Model 
\citep{cardell1977,Ben-Akiva1996multinomial, mcfadden2000mixed}. 
There are also further extensions that integrate random parameters within 
individual latent classes \citep{bujosa2010combining,greene2013revealing}.

The \gls{lcm} is developed under the assumption that individuals can be 
categorised into a set of homogeneous classes \citep{Greene2003}. 
The classes are latent in that the analyst is unable to observe which 
individual belongs to which class \citep{greene2013revealing}.  
The \gls{lcm} accounts for heterogeneity through employing different utility 
specifications for each class.
However, since the number of classes of a \gls{lcm} is finite, it only 
allows a limited number of parameter variants.
Additionally, it is difficult to determine an appropriate number of discrete 
classes in the dataset \citep{nylund2007deciding}.
In contrast, the mixed-logit model has the merit of allowing individual's 
parameters to vary randomly over a continuous distribution through simulation
\citep{Greene2003}. 

The mixed-logit model has been applied to capture a broad spectrum of 
heterogeneity sources.
Examples of applications of mixed-logit models in the literature include 
addressing diverse choice preferences 
\citep[e.g.][]{hess2005estimation,cirillo2006evidence} and
divergent willingness to pay \citep[e.g.][]{bastin2010estimating}. 
The mixed-logit model is also widely used in dealing with correlations between 
alternatives \citep[e.g.][]{brownstone2000joint} and 
across space \citep[e.g.][]{bhat2009copula}.
In general, mixed-logit models do not have a closed-form expression of the integral formula.
As such, estimation of these models relies on simulation to approximate the
mixing integration.

\subsection{Model Transferability}
\label{subsec: transferability}

Despite the fact that the theory of the mixed-logit model is clear, the practice of model 
estimation is vulnerable to errors when the sample size is small and does not comply with the high data quality requirement of the mixed-logit model estimation \citep{Greene2003}.
\emph{Model transfer} is a technique that can be used to remedy low data availability by developing a model for a new application context on the basis of a previously estimated model.
It therefore allows existing models to be of use to helping understand relatively poorly sampled areas. 
The idea of \emph{transferability} can be considered at various levels of generality.
Four typical hierarchical layers of transferability are: 
(i) underlying theory of travel behaviour; 
(ii) model structure; 
(iii) empirical specification; 
and (iv) parameter values \citep{Ben-Akiva1987, hansen1981pro, sikder2013spatial}.
In this paper we focus on the \emph{transferability} of parameter values.

Model transferability attracted intense research interest in the 1970s and 
80s \citep[e.g.][]{watson1975transferability,atherton1976transferability,galbraith1982intra,ben1979issues,lerman1981comment,louviere1981some}, 
including several practical applications \citep[e.g.][]{Barton-Aschman1981,Barton-Aschman1982,schultz1983development}.
In this period, researchers expected the parameters of travel behaviour models to remain stable in predicting travel behaviour in new contexts \citep{galbraith1982intra}.
For example, \citet{richards1975disaggregate} argue that true behavioural models should be expected to be able to make predictions for different populations and locations without adjusting model coefficients. 

Despite this optimism, later research identified that modelling in practice almost inevitably requires some adjustments to model coefficients before a model is transferred from one geographical area to another \citep{galbraith1982intra,koppelman1985geographic}.
There have been few contributions in the literature focusing on model
transferability since this period.

In the following sections, we present the mainstream model transfer methods identified by \citet{karasmaa2007evaluation}. 
To compare these methods in clear mathematical language, we use a simplified utility function as an example --- considering only the portion of utility that can be quantified by 
(i) a vector of observable attributes~$X$ and 
(ii) attributes' weights~$\Gamma_{in}$. 
In the estimation context, the individual~$n$'s utility in the choice situation~$t$ takes the following form:
\begin{equation}
    \label{eq: existingV}
    V_{in,1} = V(X_{in,1}, \Gamma_{in,1}) 
\end{equation}

\subsubsection{The Direct Application Approach}

The simplest approach to model transfer is to apply the existing model directly with no change on its parameters.
The utility in the application context takes the form
\begin{equation}
    V_{in,2} = V(X_{in,2}, \Gamma_{in,1})
\end{equation}
We refer to this as the \emph{Direct Application} approach. 

\subsubsection{The Transfer Scaling Approach}
As discussed, modelling practice finds that \emph{Direct Application} can lead to non-negligible modelling errors.
Nonetheless, it is shown that much of the \emph{transfer bias} (i.e. the error from using the coefficients from the estimation context for a new application), can be addressed through adjusting model constants and scales \citep{algers1994transferability,badoe1995comparison}.
This method represents the \emph{Transfer Scaling} approach.
It establishes that:
\begin{equation}
\label{eq: V2scale}
  V_{in,2} = \mu_{i,2} * V(X_{in,2}, \Gamma_{in,1}) + \alpha_{i,2}  
\end{equation}
where $\mu_{i,2}$ represents the scale factor for alternative~$i$ and $\alpha_{i,2} $ is an alternative specific constant adjusted for the application context.

\subsubsection{The \emph{Bayesian} Approach}
The third transfer method, the \emph{Bayesian} approach, involves re-estimating model parameters.
The transferred set of parameters is yielded through \emph{Bayesian inference}  (see function \ref{eq:bayestheorem}), instead of directly scaling the vector of parameters.
The transferred model is adapted from a conjugate prior, i.e., an existing model estimated on rich data.
Meanwhile, informative local data can be assimilated in the \emph{Bayesian} process.
In the basic \emph{Bayesian} method, the information of the application context jointly enters the \emph{Bayesian inference} as an estimated model.
In advanced \emph{Bayesian} methods, such as the \emph{Joint Context Estimation} method, samples can be assimilated one by one through iterative use of \emph{Bayesian inference}.
 
The computation of the \emph{Bayesian} approach is more complicated than the direct or transfer scaling approaches.
We omit the functions here, instead including a detailed introduction of \emph{Bayesian inference} in Section \ref{subsec: esbda simulator}.

\subsubsection{Summary}
It is a general consensus that more complex transfer approaches can better fit the application context than the \emph{direct application} approach.
A key question for the more advanced transfer methods is to what extent the difference between the coefficient values in estimation and application contexts is caused by the true behavioural differences, and how much is caused by the imprecision in parameter estimates (which is normally measured by model variance).
The \emph{transfer scaling} method uses the application context data only to correct the transfer bias, therefore the differences in sampling errors between the two datasets  are not explicitly considered \citep{karasmaa2007evaluation}.
Meanwhile, it postulates that the ratios of coefficients remain the same after transferring.
Therefore, it does not consider the difference in the weights of independent variables between the estimation population and the application population.

In contrast, the \emph{Bayesian} approach can address the problem of different sampling errors by adjusting the random variation in the utility function of the different datasets to be equal.
Furthermore, coefficients of independent variables are updated separately, therefore the transferred model can have a weighting system for the independent variables that is different to the existing model.
For these merits, the paper investigates the \emph{Bayesian} approach.

\section{Theoretical Foundations}
\label{sec: foundations}
Here we establish the theoretical foundations of the \gls{esbda} simulator. 
Firstly we provide a brief overview of the mixed-logit theory, to familiarise the reader with the notation and required background knowledge. 
Next, we provide a high-level overview of our approach, before covering the elements of the \gls{esbda} approach in detail. 

\subsection{Overview of Mixed-logit}
\label{subsec: mixed-logit}

The mixed-logit model has the capability to approximate any random utility-based \gls{dcm} to any degree of accuracy with appropriately specified distributions of the coefficients \citep{mcfadden2000mixed}.
For simplicity, the following analysis focuses on the linear utility specification and follows \citet{mcfadden2000mixed}'s notation:

In a choice situation, each alternative~$i \in {\{1,..., J\}}$ provides the decision-maker~$n$ some net benefit\slash utility.
\emph{Utility Theory} postulates that the decision-maker rationally chooses the option to maximise the utility.
As the complete utility cannot fully be observed by the modeller, it is instead modelled as the sum of the observable portion of the utility and an unknown error term.
The observable utility is a vector of observable attributes~$X_{in}$ weighted by the set of parameters~$\Gamma_{in}$. 
The parameters represent the degree of importance\slash preference that each individual assigns to each attribute \citep{Cherchi2012}.
The standard \gls{dcm} incorporates an \gls{iid} \gls{ev1} error term~$\varepsilon$ to represent unobserved utility.
Dividing the complete \emph{utility} into
the above assumptions leads to a logit model, where the \emph{utility} of choice~$i \in{J}$ is:
\begin{equation}
    U_{in} = V(X_{in}, \Gamma_{in}) + \varepsilon_{in}
\end{equation}
The corresponding probability is:
\begin{equation}
\label{eq: pynt}
    P(i|X_{in}, \Gamma_{in}) = \frac{exp\{V(X_{in}, \Gamma_{in})\}}{\sum_{j=1}^{J}{exp{\{V(X_{jn}, \Gamma_{jn})}}\}},
\end{equation}

In reality, the importance felt of each attribute~$\Gamma_{in}$ varies across population.
To model the individual heterogeneity, we allow some of the importance (parameters\slash coefficients) to take a random distribution.
Usually, modellers are interested in the ratios of certain parameters rather than their absolute value. To ensure these ratios are identifiable, some coefficients can be kept as point values.
Such specifications allow for a clear representation of the ratio between some coefficients, e.g., willingness-to-pay.
As such, $\Gamma_{in}$ can be partitioned into random parameters~$\beta'_n$ and fixed parameters~$\alpha'$.
Analogously, $X_{in}$ can be split into $z_{in}$ and $x_{in}$ pertaining to $\alpha'$ and $\beta'_n$ respectively.
Assume $\beta'_n \sim N(\zeta,  \Omega)$ where $\zeta$ denotes a mean vector and $\Omega$ represents a covariance matrix.
Accordingly, $V(X_{in}, \Gamma_{in})$ in equation \ref{eq: pynt} becomes:

\begin{equation}
    V(X_{in}, \Gamma_{in})= \alpha'z_{in} + \beta'_nx_{in}
\end{equation}
The new \emph{utility} function is function \ref{eq: utility} and probability equation \ref{eq: pynt} can be updated correspondingly.
\begin{equation}
\label{eq: utility}
U_{in} = \alpha'z_{in} + \beta'_nx_{in}+\varepsilon_{in} \\
\end{equation} 

Without a closed-form expression, estimation of mixed-logit models heavily relies on simulation tools.
The two dominant simulation means for mixed-logit models are \acrfull{msl} and the \emph{Bayesian} approach.
The principle of \gls{msl} is to search for the parameter set with which under the assumed model structure the observed data is most probable.
It is achieved by maximising a likelihood function in simulation.
In contrast, the \emph{Bayesian} approach incorporates an initial set of parameters and update the parameters over and over again using \emph{Bayesian} inference until they maintain stable values across subsequent iterations.
As the process of \emph{Bayesian} inference naturally involves iterative updates of model parameters, it provides an ideal platform for performing model transfer.

\vspace{4mm}

\subsection{Proposed Approach}
\label{subsec: esbda simulator}

In this paper, we establish  a new simulator which takes an existing model and re-estimates the parameters for a new context, exploiting both the previous parameter estimates and the new data.
Typical new contexts that could be considered are: (i) a new geographical location; (ii) a different demographic population segment; or (iii) a different time period.

\begin{figure}[!htbp]
\centering
\includegraphics[%
width=0.8\textwidth]{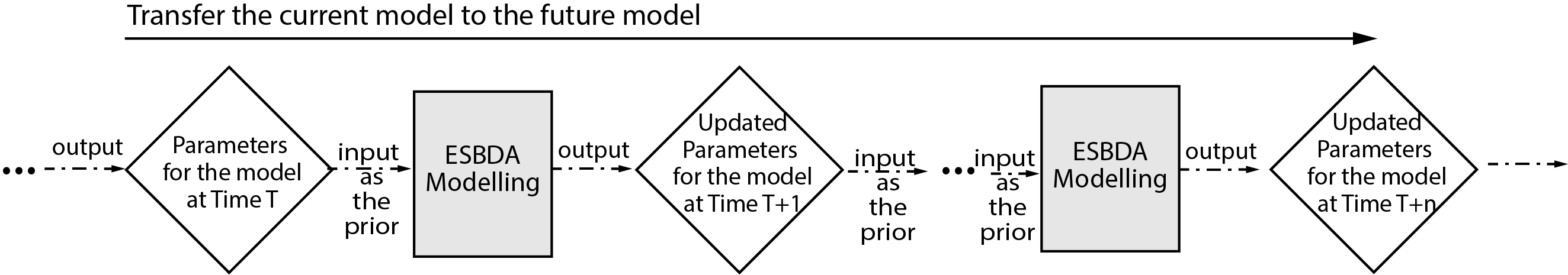}
\caption{\label{fig: update} Modelling structure for continuous update of the current model with newly available data}
\end{figure}

\begin{figure}[!htbp]
\centering
\includegraphics[%
width=0.8\textwidth]{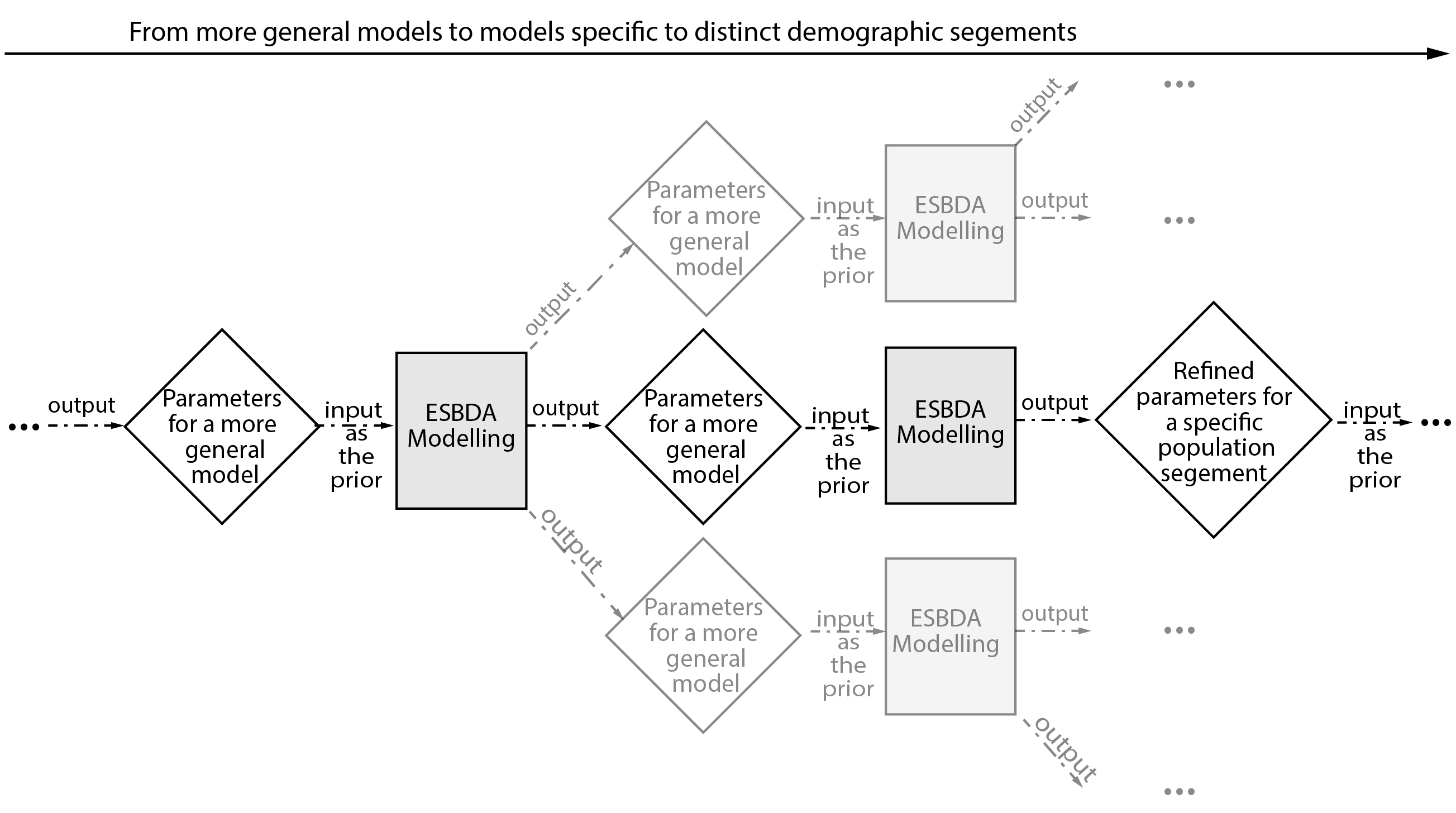}
\caption{\label{fig: hierarchy} Hierarchical modelling structure for a systematic \gls{dcm} model segmentation}
\end{figure}

A key application of this approach is updating a model for future choice behaviour estimation (see Figure \ref{fig: update}).
Another potential area of application is model segmentation, where a hierarchical modelling structure can be established to investigate heterogeneous choice behaviour hierarchically --- from a general level to specific detailed segments --- through layers of \gls{esbda} (see Figure \ref{fig: hierarchy}). 
In each level of segmentation, the coefficients, i.e., the \emph{posterior},  estimated for the upper-level model are input to the \gls{esbda} as the \emph{prior} to estimate this level \emph{posterior} coefficients, which then serve as the \emph{prior} of the next level.
\vspace{4mm}
\vspace{4mm}

The new simulator, called \acrfull{esbda}, has two central theoretical elements: \acrfull{bda} and \emph{Early Stopping}.

\subsubsection{\glsentrylong{bda}}
\label{bda review}
\gls{bda} is a term coined for time-series modelling, which describes a time-related model calibration\slash refinement technique that improves forecasting accuracy through using new information as it becomes available \citep{jazwinski1970, reich2015probabilistic}.
We extend this to a more general purpose definition as `the technique of \emph{data assimilation} through \emph{Bayesian inference} for general transfer\slash update of a previously established model.'

\emph{Bayesian inference} is the process of updating the \emph{posterior} probability~$K(\theta|Y)$ using two antecedents: a \emph{conjugate prior} probability distribution~$k(\theta)$ which reflects initial ideas about the probability, and a likelihood  function~$L(Y | \theta)$ which is derived from a statistical model for a set of the observed data~$Y = {y1,...,y_N}$.
According to Bayes theorem, we have:
\begin{equation}
\label{eq:bayestheorem}
    K(\theta|Y) = \frac{L(Y | \theta)k(\theta)}{L(Y)}
\end{equation}

We identify \emph{Bayesian inference} as a promising approach to improving 
the transferibility of mixed-logit models, because the process of updating 
the \emph{posterior} is transferable to adapting existing parameters to a new 
context.
In the context of model transfer, the previously estimated combination of 
parameters plays the role of the \emph{conjugate prior} and the combination 
of parameters to be estimated for the new context is the \emph{posterior} to 
be inferred.
Through assimilating sample data of the modelling object, the \emph{prior} 
can be transferred into a \emph{posterior} combination of parameters which 
fits to the target context.

There are two layers of \emph{Bayesian inference} applied in this study for distinct purposes.
It is important to distinguish the concept of `data assimilation', that  describes the outer layer, with the idea of `\gls{hb}', which is the process of the inner layer.
In the outer layer, the \emph{prior} and the \emph{posterior} are two versions of mixed-logit combinations of parameters and the transfer is processed through the assimilation of sample data.
We define the process of assimilating new data and updating the entire combination of parameters as `data assimilation'. 
The data assimilation process itself involves iterative internal  \emph{Bayesian inferences}, where each parameter is updated in turn on the condition of the values of the remaining parameters, in an iterative process.
The \emph{prior} and the \emph{posterior} in each iteration are single parameters (rather than the combination of all parameters).
This iterative process is the prominent `\gls{hb}', a technique introduced to estimate mixed-logit by \citet{mcculloch1994exact} and \citet{Allenby1997}
with normal distribution of coefficients, and generalised to non-normally distributed coefficients by \citet{train2001comparison}.
We will return to the detail procedures of \gls{hb} when illustrating the algorithm in Section \ref{subsec: estimation}.

\vspace{4mm}

\subsubsection{Early Stopping}

In \gls{ml}, \emph{early stopping} is a commonly used technique to terminate model training before convergence to regularise the model and prevent over-fitting \citep{Precheit1998}.
It tracks the real-time modelling error(s) on an \emph{out-of-sample} validation dataset (separate from the training data) and terminates the modelling when an \emph{early stopping} criterion is met.
This technique is not typically used in the conventional mixed-logit estimation as this model type rarely has a high dimensional parameter space and therefore has a relatively low risk of \emph{over-fitting}.

Nevertheless, we identify that \emph{early stopping} has potential benefits for mixed-logit models when working with a small sample size. 
Firstly, \emph{early stopping} can prevent the resultant model from over-fitting to the insufficient sample data which may not be representative of the population to be modelled.
In \gls{msl} estimation, the error ($E$) is automatically monitored in the form of likelihood during the \gls{msl}'s modelling progress. 
This ensures that the output parameter estimates are those that result in the lowest error ($E$).
By contrast, the training progress of traditional \emph{Bayesian} model does not examine whether or not the output parameter estimates lead to the optimal predictive error.
In this sense, while the \gls{hb} procedure uses a sampling approach and the \gls{msl} solves an optimisation problem, 
\gls{esbda} consolidates these two methods in a hybrid approach; the integration of the \emph{early stopping} procedure essentially configures the core of \gls{msl} simulator into the \emph{Bayesian} simulator, which provides a convenient way to compare the predictive error of the resulting mixed-logit model at the end of \emph{Bayesian} modelling with the smallest error during the modelling.

Furthermore, the use of an \emph{early stopping} procedure provides a good complement to the convergence sign of the mixed-logit model.
Whilst the basic \gls{hb} procedure typically terminates the simulation when the mixed-logit model converges,
the model may never converge when  sample size is small.
Under the existing estimation approach, the decision of when to stop the modelling process is arbitrary (e.g. when a default absolute number of iterations is reached). 
Instead, our use of an \emph{early stopping} procedure avoids such arbitrariness. 
Despite these potential benefits, to the best of the authors' knowledge, our effort to configure the \emph{early stopping} procedure of \gls{ml} into a mixed-logit simulator is the first. 

Among the mainstream classes of \emph{early stopping} criteria illustrated by \citet{Precheit1998}, the following criterion is the most commonly used one in \gls{ml}:
Let $E(T)$ denote the out-of-sample modelling error at epoch~$T$ and the lowest error obtained in epochs until $T$, $E_{opt}(T)$, is defined as:
\begin{equation}
    E_{opt}(T) = \min_{T^{'} \leq T}E(T^{'})
\end{equation}

It is often not the best time to halt the training immediately after the first sign of no further decrease of modelling error~($E$).
The reason is that modelling performance may hover around a plateau of no improvement or even a temporary drop before a substantial improvement.
This concern can be addressed by incorporating a delay to \emph{early stopping} with regard to the acceptable number~($k$) of epochs with no performance improvement since the last minimum $E$ occurs.
In other words, \emph{early stopping} is triggered after epoch {\small$T$}, iff $E(T) > E(T - j) \forall j \leq k$.
It is noteworthy that the output estimation is derived at epoch {\small$T - k$} rather than at {\small$T$} to retrieve the optimised model estimation.
\vspace{4mm}

\subsection{Estimation Approach of \glsentrytext{esbda}}
\label{subsec: estimation}

Our simulator is an extension to the \gls{hb} procedure for obtaining 
mixed-logit models parameters, which was initially established by 
\citet{rossi1996value} and \citet{Allenby1997}.
In this paper, we build on the algorithm introduced by \citet{Train2006}\footnote{The code is available online: \url{https://eml.berkeley.edu/Software/abstracts/train1006mxlhb.html}}.
The key extensions of the proposed 
simulator from the standard \gls{hb} procedure are on the two ends of the 
original algorithm: 
(i) the adoption of a \emph{conjugate prior} combination of parameters in the beginning 
and (ii) the \emph{early stopping} procedure to terminate modelling.

\begin{figure}[!htbp]
\centering
\includegraphics[%
width=0.8\textwidth]{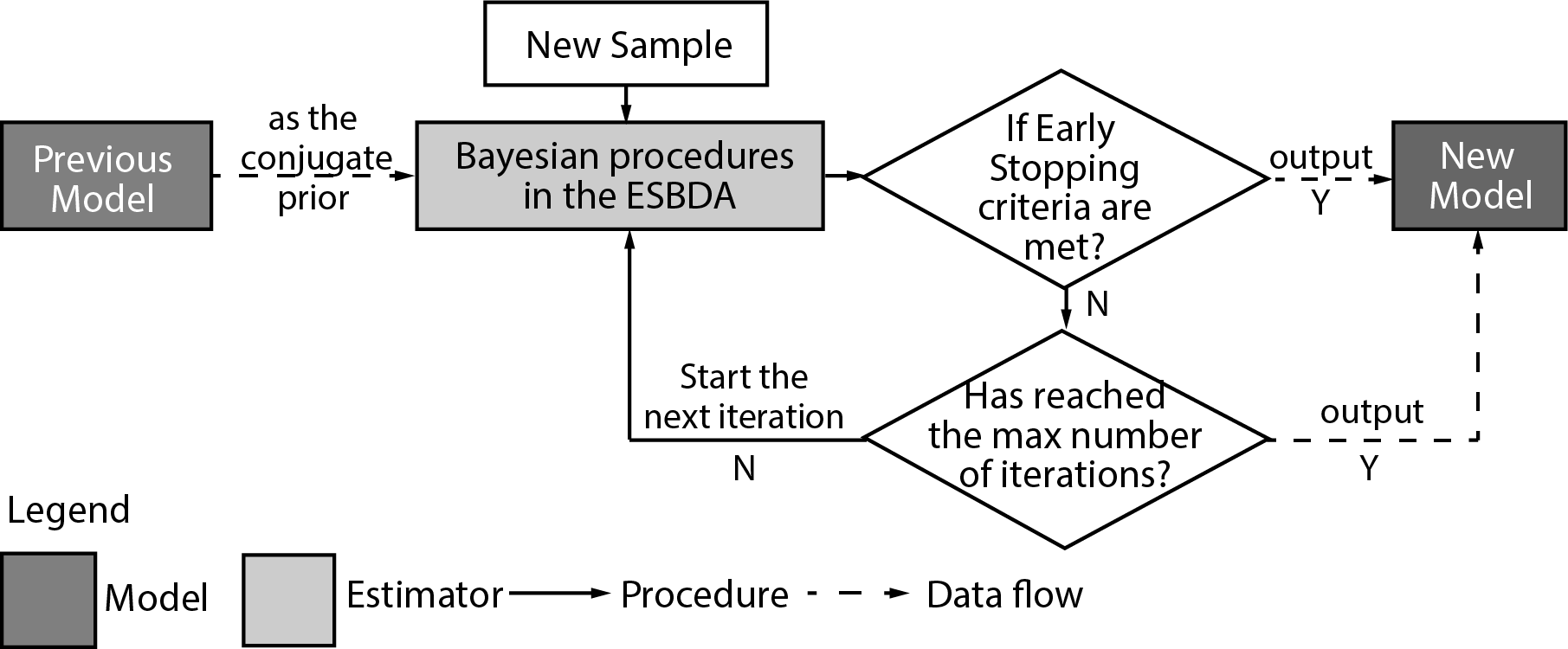}
\caption{\label{fig: esbda} Illustration of the iterative \emph{Bayes} modelling process of \gls{esbda} with an \emph{early stopping} trigger.}
\end{figure}

To present the key modelling steps more clearly, we illustrate the iterative 
\emph{Bayes} modelling process of \gls{esbda} with the 
\emph{early stopping} trigger in Figure \ref{fig: esbda}.

\gls{esbda} approximates the posterior estimates by assimilating new data 
through a \gls{mcmc} process.
The essence of \gls{mcmc} is to approximate an otherwise difficult-to-compute 
posterior distribution by draws from a Markov chain whose stationary 
distribution makes up the posterior distribution of interest 
\citep[for general treatment, see ][]{robert2013monte}.

To reserve space, we state the \gls{hb} procedure succinctly using the 
easy-to-follow multivariate normal.
For an in-detail demonstration of the \gls{hb} procedure, we direct the reader 
to Chapter 9 and 12 of \citet{train2003discrete}.

For the parameters of function \ref{eq: utility}, the conditional posteriors in each 
layer of \emph{Bayesian} inference are:
\begin{enumerate}
    \setlength\itemsep{1em}
    \item {\small$K(\beta_n | \alpha, \zeta, \Omega) \propto L(y_n | \alpha, \beta_n)\phi(\beta_n | \zeta, \Omega)$}
    \item {\small$K(\zeta | \Omega, \beta_n \forall n)$} is {\small$N(\sum_n\beta_n/N, \Omega/N))$}.
    Note {\small$\alpha$} does not enter this layer directly. Its affect on posterior {\small$\zeta$} is passed through the draws of {\small$\beta_n$ }from the first layer.
    \item {\small$K(\Omega |\zeta, \beta_n \forall n)$} is {\small$IW(K +N, (KI + N\bar{S})/(K +N))$} where {\small$\bar{S} = \sum_n(\beta_n - \zeta)(\beta_n - \zeta)^{'}/N$}. 
    Similarly, {\small$\alpha$} does not involve directly.
    \item {\small$K(\alpha | \beta_n) \propto \Pi_nL(y_n | \alpha, \beta_n)$}. 
    The \gls{mh} may be used again when the \emph{prior} on {\small$\alpha$} is essentially flat.
\end{enumerate}

The method can be conveniently adapted to variants of normal distribution simply through transformation of the underlying distribution.
We denote the weights of random \emph{utility} terms in an individual {\small$n$}'s 
\emph{utility} function as {\small$c_n$}, and {\small$c_n = T(\beta'_n)$}, 
where {\small$T$} refers to a distribution transformation which depends only 
on the latent distribution parameters and which is weakly monotonic 
(to maintain {\small$ \partial{c_{n}^{k}} / \partial{c_{n}^{\beta'_n}} \geq 0$} 
for elements in {\small$\beta_n$} or {\small$c_n$}).
The distributed random parameter is drawn in the same manner in modelling but 
it enters the \emph{utility} function in its transformed form:

\begin{equation}
\normalsize
\label{eq: UtilityTransform}
U_{in} = \alpha'z_{in}+ T(\beta'_n)x_{in}+ \varepsilon_{in}
\end{equation}

Whilst the derivation of the resulting posterior in each layer may change in other types of flexible distributions, the procedures are broadly similar.

\subsubsection{Early Stopping Parameters}
To supervise \emph{early stopping}, we employ \gls{cel}  (function \ref{eq: CEL}) 
which monitors real-time predictive performance of the estimates during modelling.
The \gls{cel} is a normalised version of the log-likelihood, whose absolute value is independent on sample size.
\begin{equation}
\normalsize
\label{eq: CEL}
\begin{split}
    G_{CEL} & = - \frac{1}{N}G_\text{log-likelihood}, \\
    & =  - \frac{1}{N}\sum_{n=1}^{N}\ln{P(i_n|x_n)}
\end{split}
\end{equation}
where $i_n$ is the index of the choice made by individual~$n$.

The threshold value of an \emph{early stopping} criterion is 
usually selected in an interactive fashion to seek the lowest generalisation 
error or yield the best `price-performance ratio'~\citep{Precheit1998}.
To guarantee model termination, the stopping criterion is complemented by a rule 
that terminates modelling after a set number of epochs.
The total number of epochs and the number of draws for simulating the 
distributed parameters are set on an ad-hoc basis.
To relieve serial correlation of \gls{mh}, draws of \emph{posterior} 
distribution of {\small$\alpha$}, {\small$\beta_n$} are retained at regular 
intervals instead of consecutively (every {\small$T_1$} epochs).
\gls{cel} is tracked and plotted every {\small$T_2$} epochs.
We will demonstrate the chosen hyperparameter values in Section \ref{subsec: parameters}.
\vspace{4mm}

\section{Experimental Methodology}
\label{sec: methodology}

Modelling experiments are carried out to benchmark the proposed 
\gls{esbda} simulator against three reference approaches. 
This section presents the set-up of the experimentation.
\subsection{Benchmarks}
\label{subsec: alternatives}
\begin{table}[!htbp]
  \centering
    \scriptsize 
    \begin{tabular}{p{6em}|p{8.5em}p{8.5em}p{8.5em}p{8.5em}}
    \hline

   Simulator & \multicolumn{2}{c}{\emph{Prior-based Bayesian} \gls{bda} simulators} &\emph{Nonconjugate prior} & Direct Application \\
   & \nohyphens{\gls{esbda}} Simulator & \gls{bda} Simulator &\emph{Bayesian} Simulator & Approach\\

     \hline
    \nohyphens{Simulation mechanism} & \emph{Bayesian} modelling & \emph{Bayesian} modelling  & \nohyphens{Bayesian modelling} & No simulation  \\
     & & & \\
    \emph{conjugate prior}  & \nohyphens{Previously estimated parameters}  & Previously estimated parameters & No-\emph{prior}  & N/A \\
     & & & \\
    \nohyphens{Early stopping procedures}  & Yes  & No    & No & N/A \\
     \hline
    \end{tabular}%
    \caption{The \gls{esbda} simulator and the benchmark simulators}
    \label{tab:simulators}
\end{table}%

The first reference approach is the Direct Application approach, i.e., the exact \emph{prior} model. 
The other reference approaches make use of Bayesian simulation. 
The \emph{Nonconjugate prior Bayesian} Simulator is adapted from the \gls{hb} procedure of \citet{Train2006}.
The parameter estimates are initiated using purely random starting values.
The estimation is therefore based solely on the limited sample data of the modelling target.
The \gls{bda} simulator initiates Bayesian estimation using the parameter values from the previously estimated model as the prior. 
It therefore exploits the potential of the \emph{conjugate prior} in \emph{Bayesian inference} and develops the new model from an informative \emph{prior} model.
The \gls{esbda} simulator then adds \emph{early stopping} to the \gls{bda} simulator,  monitoring the model performance on an out-of-sample validation set. 

\subsection{Performance Measures}
\label{subsec: simulatormeasures}
Three dimensions are investigated to assess the alternative simulators: (i) statistics of estimates, (ii) behavioural consistency of the resultant model and (iii) steadiness of the modelling process.
The monitored statistics include the statistical significance of individual parameter estimates and predictive performance of the resultant model.
In addition to \gls{cel} , which monitors the real-time predictive performance in model training (see Section \ref{subsec: estimation}), the other performance metric is \gls{gmpca} \citep{Hillel2019understanding}.
The \gls{gmpca} has a clear physical interpretation, the geometric average correctness of the output model, whilst \gls{cel} is difficult to interpret.
It is established as
\begin{equation}
    G_\text{GMPCA} = \left(\prod_{n=1}^{N}P\left(i_n|x_n\right)\right)^\frac{1}{N}
\end{equation}

The data for each experiment is divided into three sets. 
The training set is used to train the mixed-logit model, and the validation set is held to determine \emph{early stopping}.
As the validation set is used to determine the optimal parameter estimates when the out-of-sample performance is highest, it is no longer an unbiased estimate of the model's predictive power.  
As such, a separate testing set is used with the final model to test  the models' out-of-sample true predictive performance on previously unseen data.

We would like the simulator to output behaviourally-rich mixed-logit models, which
are highly interpretable and thus informative to choice behaviour studies.
Behavioural consistency of the output models is examined to eliminate estimations that are statistically significant but behaviourally meaningless.
As such, we screen sign-errors and problematic relative ratios of the output parameters. 

Finally,  the simulation progress of each modelling experiment is monitored to assess how steadily the \gls{cel} progresses throughout a complete modelling; whether the estimation results truly converge; and when \emph{early stopping} occurs.

\subsection{Case-Studies}
\label{subsec: casestudy}
The \gls{esbda} and benchmark simulators are compared on 2 case-studies: one covering vehicle purchase choice in California, and the other covering travel mode-choice in London. 
For each case-study, multiple modelling scenarios, or `levels' are considered, each using different subsamples of the data.
In each modelling scenario, the \emph{conjugate prior} for the \gls{esbda} and the \gls{bda} simulators comes from the optimal estimates at one higher level, whilst no \emph{prior} is fed to the \emph{Nonconjugate prior Bayesian} simulator. 
The Direct Transfer approach then evaluates the performance of the unmodified \emph{prior} model on the new data. 
All models are trained and evaluated on the same sample data. 

A model is first estimated for the full dataset (Level-0) using Bayesian estimation. 
Each subsequent level then represents the transfer of a model from the previous level to a smaller sample or sub-group of the population, e.g. in the Level-1 scenario, the trained model from Level-0 is transferred to a smaller sample. 
For these levels' labels, a higher number indicates a smaller sample, with less data available for model training (see Table in \ref{tab: case1level} and Table in \ref{tab: case2level}).
The process of deriving lower level models through \gls{esbda} from the Level-0 model is illustrated in Figure \ref{fig: hierarchy}.

\subsubsection{Vehicle Purchase Choice Model in California}
The first case-study models customers' willingness-to-purchase of all-electric and gas-electric vehicles using a stated preference survey conducted in California, USA. 
The baseline model for this case study is a mixed-logit model inherited from the demonstrative model of Train \& Garrett's simulator \citetext{2006}.
The model is eminently suitable for testing the alternative simulators.
The data and the \emph{utility} function specification have undergone a thorough inspection in the previous experiment (see Train \& Sonnier, 2004).
\begin{table}[!htbp]
  \centering
  \scriptsize 
    \begin{tabular}{p{18em}p{5em}p{5em}p{5em}}
    \toprule
    Variable & Symbol & Coefficient & Distribution \\
    \midrule
    Purchase price ($*10,000$) & $p_{in}$ & $\beta_{\text{price},n}$     & lognormal \\
    Operating cost (\$/month)  & $o_{in}$ & $\beta_{\text{operate},n}$      & lognormal \\
    Range in hundreds of miles (0 if not electric)  & $r_{in}$ & $\beta_{\text{range},n}$      & lognormal \\
    Electric (boolean)  & $e_{in}$ & $\beta_{\text{electric},n}$     & normal \\
    Hybrid (boolean)  & $h_{in}$ & $\beta_{\text{hybrid},n}$      & normal \\
    High performance (boolean)  & $s_{in}$ & $\alpha_\text{high}$      & fixed \\
    Medium or high performance (boolean)  & $m_{in}$ & $\alpha_\text{midhigh}$      & fixed \\
    \bottomrule
    \end{tabular}%
    \caption{Variables and coefficients of the California vehicle purchase choice model, and their distribution types}
  \label{tab: coefficient1}
\end{table}%

In this case-study, each respondent faces 15 rounds of choices, with three options given in each choice.
The sample data contain respondents' stated choices and characteristics of the alternative vehicles, such as engine type (i.e., electric, hybrid or gasoline), purchase price, etc.
As we do not have socio-economic details for survey respondents, the samples for Levels 1 and 2 are randomly selected from the full dataset. 
This means that the expected values of the true model parameters do not change for different modelling levels. 
As such, this case-study investigates the potential for overfitting on small training sample sizes for each simulator.

\begin{table}[!htbp]
  \centering
   \scriptsize 
    \begin{tabular}{lp{18em}lll}
    \toprule
    \multicolumn{1}{l}{\multirow{1}[0]{*}{ Size level}} & \multirow{1}[0]{*} {Modelling object} & \multicolumn{3}{c}{Sample size} \\
       & \multicolumn{1}{c}{} & \multicolumn{1}{l}{Training} & \multicolumn{1}{l}{Validation} &  \multicolumn{1}{l}{Test}\\
    \midrule
   Level-0  &  Full dataset (100 respondents). & 1484 & - & - \\
    & & & & \\
   Level-1  & Randomly selected sample: 20 individuals per set& 300 & 300 & 300\\
    & & & & \\
   Level-2  & Randomly selected sample: 5 individuals per set& 75 & 75 & 75 \\ 
    \bottomrule
    \end{tabular}%
  \caption{Levels of sample size of the California vehicle purchase choice model}
   \label{tab: case1level}
\begin{tablenotes}
      \scriptsize
      \item 
      \begin{flushleft}
      \emph{Note.} 
    The sample size of the full dataset is not 1500 because some choice records  are missing.
      \end{flushleft}
    \end{tablenotes}
\end{table}%
The model incorporates uniform weights for the same variables across the alternative choices.
The uniform \emph{utility} function of alternative~$i$ is as follows.
The explanatory variables and the coefficients of the model are presented in Table \ref{tab: coefficient1}.
\begin{equation}
\label{eq: ucase1}
  \normalsize
\begin{split}
    U_{in} =& (\mu_{\text{price}}+\sigma_{\text{price}} \zeta_{\text{price},n})p_{in} + (\mu_{\text{operate}}+\sigma_{\text{operate}} \zeta_{\text{operate},n})o_{in} + \\ &(\mu_{\text{range}}+\sigma_{\text{range}} \zeta_{\text{range},n})r_{in} + (\mu_{\text{electric}}+\sigma_{\text{electric}} \zeta_{\text{electric},n})e_{in} +  \\
    &(\mu_{\text{hybrid}}+\sigma_{\text{hybrid}} \zeta_{\text{hybrid},n})h_{in} + \alpha_{\text{high}}s_{in} +\alpha_{\text{midhigh}}m_{in} + \varepsilon_{in}
\end{split}
\end{equation}

\subsubsection{Travel Mode Choice Model in London}

The second case-study  considers passenger mode-choice using revealed preference data collected in London, UK.
The dataset, available online, is adapted from a closely tailored London travel dataset\footnote{available on \url{https://www.icevirtuallibrary.com/doi/suppl/10.1680/jsmic.17.00018}.} 
which recreates the travel mode choice-set that are faced by the respondent at the time of travel \citep{hillel2018recreating}.

The ratio of parameters for time and cost, known as the \emph{Value of Time}, is of particular interest in transport modelling.
To ensure this ratio is well defined, we assign a normal distribution to the coefficient for cost and maintain all other coefficients\slash constants as fixed-values.
People's perception and valuation of time vary when travelling in different modes.
To reflect this, we set alternative-specific parameters for the \emph{utility} functions of the four modes, i.e., driving ({\small$U_{\text{driving},n}$}), public transit ({\small$U_{\text{public},n}$)}, cycling ({\small$U_{\text{cycling},n}$}), and walking ({\small$U_{\text{walking},n}$}). 
The \emph{utility} functions are as follows (functions \ref{eq: ud}--\ref{eq: uw})\footnote{To conserve space, we use $\beta_{\text{cost},n}$ to denote $(\mu_\text{cost}+\sigma_\text{cost} \zeta_{\text{cost},n})$ in the rest of this paper.}.
\begin{align}
    \normalsize
    \label{eq: ud}
    \normalsize
    U_{\text{driving},n} =~& (\mu_\text{cost}+\sigma_\text{cost} \zeta_{\text{cost},n})c_{n\_d} + \alpha _\text{driving-time}t_{d,n} + \alpha_{\text{var}}\nu_{d,n} + \varepsilon_{\text{driving},n} \\
    \label{eq: up}
    \nonumber
    U_{\text{public},n} =~& (\mu_\text{cost}+\sigma_\text{cost} \zeta_{\text{cost},n})c_{p,n}+\alpha_{\text{access-time}}t_{a,n} +\alpha_{\text{bus-time}}t_{b,n}+ \\
    &\alpha_{\text{rail-time}}t_{r,n}+
    \alpha_{\text{change-walking-time}}t_{\text{change1},n}+\alpha_{\text{change-waiting-time}}t_{\text{change2},n}+\\
    \nonumber
    &C_\text{public}+\varepsilon_{\text{public},n} \\
    \label{eq: uc}
    \normalsize
    U_{\text{cycling},n} =~& \alpha_{\text{cycling-time}}t_{c,n} + C_\text{cycling} + \varepsilon_{\text{cycling},n} \\
    \label{eq: uw}
    \normalsize
    U_{\text{walking},n} =~& \alpha_{\text{walking-time}}t_{w,n} + C_\text{walking} \varepsilon_{\text{walking},n}
\end{align}

For the London model, we have detailed socioeconomic information for each individual in the dataset.
This provides a platform to illustrate model transfer for hierarchical segmentation (as Figure \ref{fig: hierarchy} illustrates).
Levels 1 to 3 represent data for specific population segments taken from the original sample. Since each modelling level represents a different demographic group, each modelling level represents a true transfer to a new application.
As such, the expected values of the true model parameters are different for each modelling level.
This case-study therefore investigates the ability of each simulator to transfer a model to a new application. 

Following the recommendations in a systematic review of classification methodologies  \citep{hillel2020systematic}, the validation and test folds are sampled grouped by household to ensure that trips of the same household are not classified into different folds.
At each level, the training, validation and test datesets are allocated the same number of households. 
Therefore, the selected the number of households in each fold is fixed while the number of trips may vary.
\begin{table}[!htbp]
  \centering
   \scriptsize 
    \begin{tabular}{p{14em}p{6em}p{7.5em}p{4.5em}}
    \toprule
    Variable\slash Constant & Symbol & Coefficient & Distribution\\
    \midrule
    Driving cost & $c_{d,n} $ & $\beta_{\text{cost},n}$ & normal \\
    Public transport cost & $c_{p,n} $ & $\beta_{\text{cost},n}$ & normal \\
    Driving time  & $t_{d,n}$ & $\alpha_{\text{driving-time}}$     &  fixed \\
    Access time  & $t_{a,n}$ & $\alpha_{\text{access-time}}$     & fixed \\
    In-vehicle time on bus  & $t_{b,n}$  & $\alpha_{\text{bus-time}}$     & fixed \\
    In-vehicle time on rail  & $t_{r,n}$ & $\alpha_{\text{rail-time}}$     &  fixed \\
    Interchange walking time  & $t_{\text{change1},n}$   & $\alpha_{\text{change-walking-time}}$     & fixed \\
    Interchange waiting time  & $t_{\text{change2},n}$  & $\alpha_{\text{change-waiting-time}}$     & fixed \\
    Cycling time  & $t_{c,n}$ & $\alpha_{\text{cycling-time}}$     &fixed \\
    Walking time  & $t_{w,n}$  & $\alpha_{\text{walking-time}}$     &  fixed \\
    Traffic variability  & $\nu_{d,n} $   &  $\alpha_\text{traffic}$   & fixed \\
    Constant of the Public transit mode  & -  & $C_\text{public}$   & fixed \\
    Constant of the Cycling mode  & -   & $C_\text{cycling}$ &   fixed \\
    Constant of the Walking mode  & -  & $C_\text{walking}$  & fixed \\
    \bottomrule
    \end{tabular}%
  \caption{Variables and Coefficients of the London Travel Mode Choice Model, and the distributions of the coefficients}
   \label{tab: coefficient2}
\end{table}%

\begin{table}[!htbp]
  \centering
   \scriptsize 
    \begin{tabular}{lp{17em}lll}
    \toprule
    \multicolumn{1}{l}{\multirow{1}[0]{*}{ Sample size}} & \multirow{1}[0]{*} {Modelling object} & \multicolumn{3}{l}{Sample size} \\
       & \multicolumn{1}{c}{} & \multicolumn{1}{p{4em}}{Training} & \multicolumn{1}{p{5em}}{Validation} &  \multicolumn{1}{p{5em}}{Test}\\
    \midrule
   Level-0  &  All journeys, regardless of travel purpose time period of travelling or the traveller's attributes, income, age, etc. & 8331 & - & - \\
    & & & &\\
   Level-1  &  General home-office journeys, regardless of time period of travelling or the traveller's attributes, income, age, etc. & 613 & 735 & 643 \\
    & & & &\\
   Level-2  & Home-office journeys during morning peak-time\footnotetext{Transport for London charges higher fares at Peak Time on workdays. Peak fares are applied during 06:30-09:30, and between 16:00 to 19:00}, regardless of the traveller's attributes, income, age, etc. & 266 & 264 & 325\\
    & & & &\\
   Level-3  & Home-office journeys during morning peak-time; the 26-35-year-old people whose household income is between £25,000-£49,999. & 26 & 27 & 36\\    
    \bottomrule
    \end{tabular}%
  \caption{Levels of sample size of the London travel mode choice model, and the corresponding modelling objective at each level}
   \label{tab: case2level}
    \begin{tablenotes}
      \scriptsize
      \item 
      \begin{flushleft}
      \emph{Note.} 
     While the training,validation and test datesets are allocated the same number of households at each level, the number of trips varies among individuals. As such, the number of trips in each fold varies.
      \end{flushleft}
    \end{tablenotes}
\end{table}%

\subsection{Chosen Hyperparameter Values}
\label{subsec: parameters}
We set $T_1 = 10$ as the interval between checkpoints for tracking \gls{cel} and $T_2 = 20$ for plotting \gls{cel} draws.
For \emph{early stopping}, the maximum number of epochs that we allow no performance improvement is $k=200$.
As a complementary criterion, the modelling would be terminated after an absolute value of 10,000 epochs if \emph{early stopping} does not occur.

\section{Results}
\label{sec: results}
In this section, we investigate the experimental results with a particular focus on the small-sample properties of the alternative simulators.

\subsection{Interpreting the Results}
\label{subsec: interpret results}
\begin{figure}[!htbp]
\centering
\includegraphics[%
width=\textwidth]{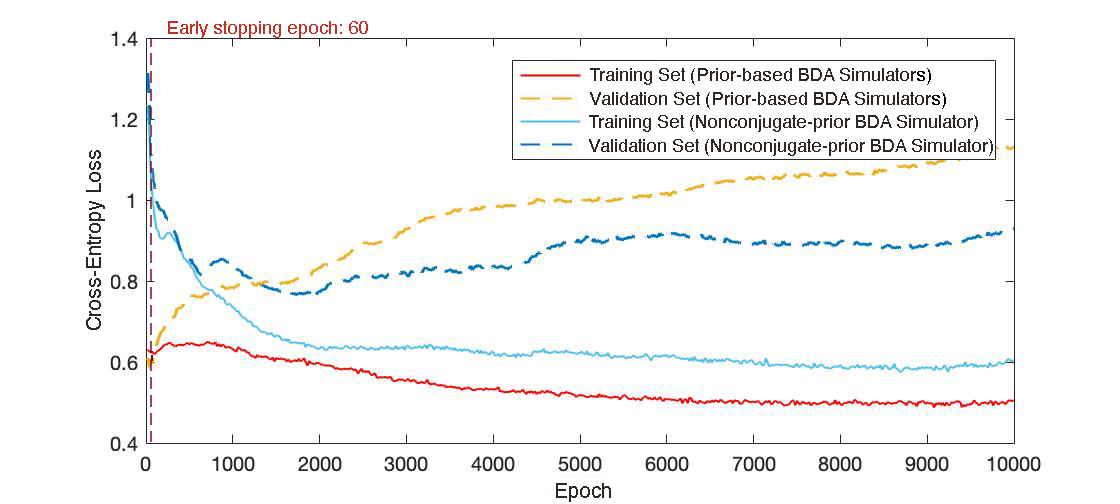}
\caption{\label{fig: Case2level3} Comparison of \acrfull{cel} of the \emph{conjugate-prior}-based \gls{bda} and the \emph{Nonconjugate-prior Bayesian} Simulator (the London model, Level-3)}
\end{figure}

Performance of alternative simulators is compared on the grounds of (i) statistics of the output parameters combination,
(ii) behavioural consistency of the output mixed-logit models to an empirical model (see Table \ref{tab: Case1level0} to \ref{tab: Case2level3});
and (iii) the steadiness of the simulators' modelling progress (Figure \ref{fig: Case1level1} to \ref{fig: Case2level3}).

In the tables, we highlight statistical insignificance, sign-errors as well as parameter estimates which may be statistically significant but are highly inconsistent with empirical modelling results.
The investigation of behavioural consistency is mainly based on the judgement of the ratios of other parameters against the coefficient for cost (which we call monetary ratio in the paper).
We highlight in the tables where the monetary ratio deviates over two orders of magnitude from the highest performing model from the level above.
In the London example, the travel time-cost ratios (i.e. Value of Time) are of our particular research interest.

For each modelling level, the convergence progresses of alternative simulators are plotted on a graph (e.g. Figure  \ref{fig: Case1level1}). 
The difference between the \gls{esbda} and \gls{bda} simulators lies only in \emph{early stopping} and therefore they have the same \gls{cel} curve until \emph{early stopping} occurs.
They are both noted as prior-based \gls{bda} simulators and are represented by the same (red and yellow) lines in the plots.
Epoch~$ (T - k) $ when \gls{esbda} outputs its estimation is marked by a red vertical dashed line. 
The \gls{cel} of each simualtor is tracked for 10000 epochs. 
Whilst the training of \gls{esbda} is terminated at epoch~$T$ if \emph{early stopping} occurs, the \glspl{cel} curves of \gls{bda} after \gls{esbda}'s \emph{early stopping} illustrates how \glspl{cel} would change if \emph{early stopping} did not occur.

\subsection{Illustrating the Results}
\label{subsec: results}
The change of the alternative simulators' performance from Level-0  to lower level modelling scenarios follows the same trend in two case studies when the sample size reduces.
To avoid repetitive analysis, we present the experimental results in order of levels rather than by case studies. 

\subsubsection{Level-0}
  
\begin{table}[!htbp]
\centering
\scriptsize
\begin{tabular}{l|ll|ll}
\toprule
Simulator  & \multicolumn{2}{p{12em}|}{\emph{Bayesian} Simulators  \newline (including
\gls{esbda}, \gls{bda} \& \emph{Nonconjugate-prior Bayesian} Simulator)} & \multicolumn{2}{p{12em}}{Direct Application Approach} \\
\hline
\multicolumn{5}{c}{Random coefficient}   \\
Latent  & & & &  \\
   & Mean              & StDv            & Mean             & StDv                     \\
$\mu_\text{price}$   & -0.9166{\textsuperscript{***}} & 0.1851 &  -0.9056 & 0.1850 \\
$\sigma_\text{price}$ & 1.6340{\textsuperscript{***}}  & 0.3027 & 1.5539{\textsuperscript{***}}  & 0.4392 \\
$\mu_\text{operate}$   & -5.2851{\textsuperscript{***}} & 0.4734 & -5.2620{\textsuperscript{***}} & 0.5464 \\
$\sigma_\text{operate}$ &3.1558{\textsuperscript{***}}  & 0.9111 & 3.5568  & 1.4296 \\
$\mu_\text{range}$  &-1.6768{\textsuperscript{***}} & 0.4531 & -1.7435{\textsuperscript{**}} & 0.6222 \\
$\sigma_\text{range}$                         & 1.7364{\textsuperscript{**}}  & 0.5399 & 1.5896  & 0.8609 \\
$\mu_\text{electric}$              &        -1.3494{\textsuperscript{***}} & 0.2456 &  -1.4121{\textsuperscript{***}} & 0.3131 \\
$\sigma_\text{electric}$                      &1.8121{\textsuperscript{***}}  & 0.4348 &  2.0655{\textsuperscript{**}}  & 0.7320 \\
$\mu_\text{hybrid}$                           &0.7314{\textsuperscript{***}}  & 0.2200 &  0.6801{\textsuperscript{***}}  & 0.2059 \\
$\sigma_\text{hybrid}$                           &1.9992{\textsuperscript{***}}  & 0.3606 &  1.8305{\textsuperscript{**}}  & 0.5583\\
Simulated  & & & &   \\
$\beta_{\text{price},n}$              & -0.8663  & 1.3747 & -0.8215  & 1.3838 \\
$\beta_{\text{operate},n}$             & -0.0244  & 0.0844 & -0.0296  & 0.0822 \\
$\beta_{\text{range},n}$   & 0.4988  & 0.9310 & 0.4137  & 0.8507 \\
$\beta_{\text{electric},n}$             &-1.2958 & 1.3916 & -1.4304 & 1.4424 \\
$\beta_{\text{hybrid},n}$                   & 0.7978  & 1.4029 &  0.6838  & 1.4024 \\
\hline
\multicolumn{5}{c}{Fixed coefficient} \\
$\alpha_\text{high}$       & 0.1025 & 0.0980 & 0.1058 & 0.1005 \\
$\alpha_\text{midhigh}$      & 0.5729{\textsuperscript{***}} & 0.1021 & 0.5763{\textsuperscript{***}} & 0.1033 \\
\bottomrule
\end{tabular}
\caption{Modelling Estimates (the California model, Level-0)}
\label{tab: Case1level0}
\begin{tablenotes}
      \scriptsize
      \item 
      \begin{flushleft}
      \emph{Note.} 
     The three \emph{Bayesian} simulators, the \emph{Nonconjugate-prior Bayesian}, \gls{bda} and \gls{esbda} output identical estimates as there is no conjugate-prior model input to  \gls{bda} or \gls{esbda} at Level-0 and \emph{early stopping} does not occur to \gls{esbda}.
      \newline \textsuperscript{*} $p <0.05$; \textsuperscript{**} $p <0.01$; \textsuperscript{***} $p <0.001$
      \end{flushleft}
    \end{tablenotes}
\end{table}

\begin{table}[!htbp]
\centering
\scriptsize 

\begin{tabular}{p{8em}|p{8em}p{8em}}
\hline
Simulator  & \multicolumn{2}{p{18em}}{\emph{Bayesian} Simulators  \newline (including
\gls{esbda}, \gls{bda} \& \emph{Nonconjugate-prior Bayesian} Simulator)}  \\ \hline
\multicolumn{3}{c}{Random coefficient}   \\
Latent  & &   \\
   & Mean              & StDv                  \\
$\mu_\text{cost}$ & -0.1571\textsuperscript{***}  & 0.0317 \\
$\sigma_\text{cost}$ & 0.0174   & 0.0251  \\
Simulated & &  \\
$\beta_{\text{cost},n}$              & -0.1605\textsuperscript{*} & 0.0116 \\
\hline
\multicolumn{3}{c}{Fixed coefficient}  \\
$\alpha_{\text{driving-time}}$       & -3.4996\textsuperscript{**}   & 1.4951  \\
$\alpha_{\text{access-time}}$         & -3.4173\textsuperscript{***}   & 0.7521  \\
$\alpha_{\text{bus-time}}$     & -2.2110\textsuperscript{*}   & 1.0712 \\
$\alpha_{\text{rail-time}}$        & -2.3821\textsuperscript{**}   & 0.6132  \\
$\alpha_{\text{change-walking-time}}$        & -1.9474\textsuperscript{**}   & 0.5913  \\
$\alpha_{\text{change-waiting-time}}$       & -2.6313\textsuperscript{***}   & 0.4553 \\
$\alpha_{\text{cycling-time}}$      & -4.6405\textsuperscript{***}   & 1.0172 \\
$\alpha_{\text{walking-time}}$       &-6.2339\textsuperscript{***}   & 0.6315 \\
$\alpha_\text{traffic}$      & -5.1859\textsuperscript{***}   & 1.0467 \\
$C_\text{public}$        & 1.7403   & 1.4935  \\
$C_\text{cycling}$      & 0.2730   & 0.1501  \\
$C_\text{walking}$       & 3.5505\textsuperscript{***}    & 0.4235\\
\bottomrule
\end{tabular}
\caption{Modelling estimates (the London model, Level-0)}
\label{tab: Case2level0}
\begin{tablenotes}
      \scriptsize
      \item 
      \begin{flushleft}
      \emph{Note.} As there is no prior model in the Level-0 experiment and early stopping does not occur to \gls{esbda}, the three \emph{Bayesian} simulators generate identical estimation. 
      \newline The Direct Application approach is not applicable since there is no empirical model readily usable to the London model.
      \newline
      \textsuperscript{*} $p <0.05$; \textsuperscript{**} $p <0.01$; \textsuperscript{***} $p <0.001$;   Behavioural inconsistent parameter\slash price ratio: \textcolor{blue}{\textsuperscript{!}}: 2 orders of magnitude deviation from the prior model; \textcolor{blue}{\textsuperscript{!!}}: 3 orders of magnitude deviation.
      \end{flushleft}
    \end{tablenotes}
\end{table}
As explained in Section \ref{subsec: casestudy}, the primary purpose of the Level-0 modelling is to derive a `mother model' to feed \emph{conjugate prior} parameters to the lower level modelling, rather than to benchmark alternative estimation approaches.
No prior model is used in the highest level (Level-0) experiment and \emph{early stopping} does not occur, given the large training samples. As such, the three \emph{Bayesian} simulators output identical parameter estimates.
Therefore, to reserve space, Level-0 plots are omitted, and the identical estimates of the three \emph{Bayesian} simulators are presented by a single column in Table \ref{tab: Case1level0} and \ref{tab: Case2level0}.

Whilst there is no empirical model readily usable to the Level-0 London experiment,
the Direct Application approach is applicable to the Level-0 California Model as we use exactly the same \emph{utility} function and dataset with Train \& Garrett's demonstrative modelling.
In Table \ref{tab: Case1level0}, we compare the estimation of the \emph{Bayesian} simulators to that of the Direct Application approach, which applies the original algorithm of Train \& Garrett's.
As the table shows, the \emph{Bayesian} simulators' output parameters are highly consistent to Train \& Garrett's estimation.
The significant consistency to the literature results demonstrates that the \emph{Bayesian} simulators developed by us are at least functional estimators for mixed-logit models.

\subsubsection{Level-1}
\begin{figure}[!htbp]
\centering
\includegraphics[%
width=\textwidth]{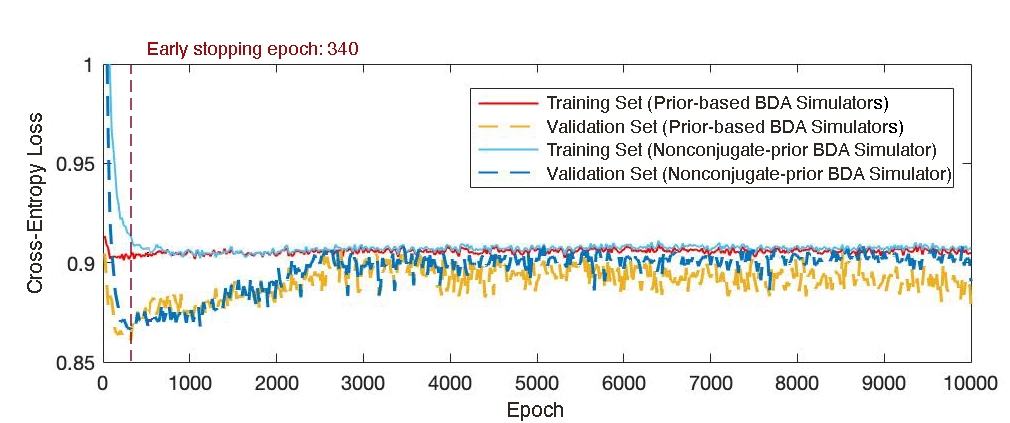}
\caption{\label{fig: Case1level1} Comparison of \acrfull{cel} of the \emph{conjugate-prior}-based \gls{bda} and the \emph{nonconjugate-prior} \emph{Bayesian} Simulator (the California model, Level-1) \newline \scriptsize{The vertical red dashed line marks epoch~$(T - k)$ when \gls{esbda}'s estimation is output.}}
\end{figure}

For Level-1, the training sample of the California model is reduced to 300 choices made by 20 individuals.
For every simulator based on each dataset, the \gls{cel} reaches a relatively stable asymptote.
There is only a slight difference in the stable training set \gls{cel} levels between the \emph{conjugate-prior} and the \emph{nonconjugate-prior} simulators.
While the training set \gls{cel} curves are relatively smooth (Figure \ref{fig: Case1level1}), there are noticeable fluctuations in the validation set \gls{cel}.
Specifically, the validation set \gls{cel} value can quickly jump by 2\% within merely 20 epochs.
Under the fluctuation, \gls{esbda} is terminated by \emph{early stopping} after 540 epochs and the optimal modelling results are outputs at the 340\textsuperscript{th} epoch.
As Table \ref{tab: Case1level1} shows, the model with the best predictive statistics as well as the best behavioural consistency is estimated by the \gls{esbda} simulator.
In contrast, the monetary ratios of a few parameter estimates by the two reference simulators diverge far from the ratios of the prior model to an extent that the estimation results are considered questionable from a behavioural perspective.
\begin{table}[!htbp]
\centering
\scriptsize
\begin{tabular}{p{6.5em}|ll|ll|ll|ll}
\toprule
Simulator  & \multicolumn{2}{p{8.5em}|}{\gls{esbda} Simulator (at the 340\textsuperscript{th} epoch)} & \multicolumn{2}{p{8.5em}|}{\gls{bda} Simulator} & \multicolumn{2}{p{8.5em}|}{\emph{Nonconjugate-prior Bayesian} Simulator} & \multicolumn{2}{p{8.5em}}{Direct Application Approach}  \\
\hline
\multicolumn{9}{c}{Random coefficient}\\
Latent  & & & & & & & \\
   & Mean              & StDv            & Mean             & StDv           & Mean                & StDv           & Mean                & StDv                  \\
$\mu_\text{price}$                                              & -1.5132\textsuperscript{***} & 0.3949 & -1.8227\textsuperscript{*}  & 0.7519 & -1.7633\textsuperscript{**}  & 0.6631 & -0.9166 & 0.1851  \\
$\sigma_\text{price}$                                           & 1.8336\textsuperscript{**}  & 0.6433 & 3.0558   & 2.7459 & 2.7880    & 2.1482 & 0.7751 & 0.3027\\
$\mu_\text{operate}$                                             & -5.8014\textsuperscript{***}  & 0.7751   & -7.0237\textsuperscript{***}  & 1.9313 & -10.9885\textsuperscript{**} & 3.4405 & -5.2851 & 0.4734\\
$\sigma_\text{operate}$                                          & 3.1118\textsuperscript{**}  & 0.9784 & 3.5396  & 3.1166 & 5.3340    & 7.1342 & 3.1558 & 0.9111\\
$\mu_\text{range}$                                              & -2.4011\textsuperscript{**}  & 0.7858  & -10.2657\textsuperscript{*} & 5.0484 & -4.5737  & 2.2737 &-1.6768 & 0.4531 \\
$\sigma_\text{range}$                                           & 2.1558\textsuperscript{*}  & 0.8384 & 4.0078   & 4.4142 & 2.3872   & 1.7787  & 1.7364 & 0.5399 \\
$\mu_\text{electric}$                                           & -1.5876\textsuperscript{***} & 0.3424  & -1.4709\textsuperscript{**}  & 0.5043 & -1.4075\textsuperscript{**}  & 0.5076 &        -1.3494 & 0.2456\\
$\sigma_\text{electric}$                                        & 1.4239\textsuperscript{*}   & 0.5577 & 2.0086   & 1.3521 & 2.0442   & 1.2034 &1.8121 & 0.4348 \\
$\mu_\text{hybrid}$                                             & 0.5282   & 04048 & 0.4076   & 0.4216 & 0.5020    & 0.3756 &0.7314  & 0.2200\\
$\sigma_\text{hybrid}$                                          & 1.2361\textsuperscript{*}  & 0.5409 & 1.4418   & 0.7715 & 1.4423   & 0.7541  &1.9992 & 0.3606\\        
\hline
Simulated & & & & & &  \\
$\beta_{\text{price},n}$                                        & -0.5221  & 0.8996  & -0.6549   & 1.6718 & -0.6193   & 1.4678  & -0.8663  & 1.3747\\
$\beta_{\text{operate},n}$                                      & -0.0138  & 0.0513 & -0.0049   & 0.0215 &\textcolor{blue}{-0.0002}\textsuperscript{\textcolor{blue}{!!}}    & 0.0014 & -0.0244  & 0.0844\\
$\beta_{\text{range},n}$                                        & 0.2797  & 0.6189 & \textcolor{blue}{0.0002}\textsuperscript{\textcolor{blue}{!}}   & 0.0007 & 0.0349   & 0.0667 & 0.4988  & 0.9310\\
$\beta_{\text{electric},n}$                                     & -1.5761 & 1.2672 & -1.4330   & 1.4882 & -1.3701  & 1.5028 &-1.2958 & 1.3916\\
$\beta_{\text{hybrid},n}$                                       & 0.5383  & 1.1049 & 0.4356   & 1.2176 & 0.5344   & 1.2181  & 0.7978  & 1.4029\\
\hline
\multicolumn{9}{c}{Fixed coefficient} \\
$\alpha_\text{high}$                                            & 0.0958  & 0.1783 & 0.0866   & 0.1965 & 0.0883   & 0.2016 & 0.1025 & 0.0980 \\
$\alpha_\text{midhigh}$                                         & 0.8014\textsuperscript{***}  & 0.1928 & 0.8682\textsuperscript{***}   & 0.2178 & 0.8588\textsuperscript{***}   & 0.2201 & 0.5729 & 0.1021 \\
\hline
\multicolumn{9}{c}{Modelling error } \\
 & CEL & GMPCA & CEL & GMPCA & CEL & GMPCA & CEL & GMPCA\\
Validation set & 0.8695 & 0.4192 &0.8788 & 0.4153 & 0.8906 & 0.4104 & 0.9124 & 0.4015\\
Test set & 1.1683 & 0.3109 & 1.2283 & 0.2928 & 1.2294 & 0.2925 & 1.2475 & 0.2972\\ 
\bottomrule
\end{tabular}
\begin{tablenotes}
      \scriptsize
      \item 
      \begin{flushleft}
      \emph{Note.} \textsuperscript{*} $p <0.05$; \textsuperscript{**} $p <0.01$; \textsuperscript{***} $p <0.001$;  \newline Behavioural inconsistent parameter\slash price ratio: \textcolor{blue}{\textsuperscript{!}}: 2 orders of magnitude deviation from the prior model; \textcolor{blue}{\textsuperscript{!!}}: 3 orders of magnitude deviation.
      \end{flushleft}
    \end{tablenotes}
\caption{Modelling estimates (the California model, Level-1)}
\label{tab: Case1level1}
\end{table}
\begin{table}[!htbp]
\centering
\scriptsize 
\begin{tabular}{p{6.5em}|ll|ll|ll|ll}
\hline
Simulator  & \multicolumn{2}{p{8.5em}|}{\gls{esbda} Simulator (no \emph{early stopping})} & \multicolumn{2}{p{8.5em}|}{\gls{bda} Simulator} & \multicolumn{2}{p{8.5em}|}{\emph{Nonconjugate-prior Bayesian} Simulator} & \multicolumn{2}{p{8.5em}}{Direct Application Approach}  \\
              \hline
\multicolumn{9}{c}{Random coefficient}   \\
Latent  & & & & & &  \\
   & Mean              & StDv            & Mean             & StDv           & Mean                & StDv    & Mean                & StDv
   \\
$\mu_{\text{cost}}$ & -0.1531\textsuperscript{***}  & 0.0409 & -0.1531\textsuperscript{***}  & 0.0409 & -0.1706\textsuperscript{***}  &  0.0144 & -0.5171 & 0.0317\\
$\sigma_{\text{cost}}$ & 0.0149   & 0.0275 & 0.0149   & 0.0275 & 0.0147  & 0.0076 & 0.0174 & 0.0251 \\
Simulated & & & & & &  \\
$\beta_{\text{cost},n}$              & -0.1641  & 0.2182 & -0.1641  & 0.2182 & -0.1816 & -0.1202  & -0.1605& 0.0116 \\
\hline
\multicolumn{9}{c}{Fixed coefficient}  \\
$\alpha_{\text{driving-time}}$       &-4.2537\textsuperscript{**}   & 1.5895 &-4.2537\textsuperscript{**}  & 1.5895 & -0.6409 &  0.8530  & -3.4996   & 1.4951\\
$\alpha_{\text{access-time}}$         & -5.7911\textsuperscript{***}   & 1.2652 & -5.7911\textsuperscript{***}  & 1.2652 & -4.1264\textsuperscript{*}  &  1.8979  & -3.4173  & 0.7521\\
$\alpha_{\text{bus-time}}$     & -3.4614\textsuperscript{**}   & 1.0902 & -3.4614\textsuperscript{**}  & 1.0902 & -1.6666 &  0.9127 & -2.2110   & 1.0712\\
$\alpha_{\text{rail-time}}$        & -4.2861\textsuperscript{*}   & 1.6784 & -4.2861\textsuperscript{*}   & 1.6784 & -0.4644 &  1.2405  & -2.3821 & 0.6132\\
$\alpha_{\text{change-walking-time}}$        & -4.8911\textsuperscript{*}   & 2.4391 & -4.8911\textsuperscript{*}  & 2.4391 & -0.2100 & 0.6514 & -1.9474  & 0.5913\\
$\alpha_{\text{change-waiting-time}}$       & -4.0699\textsuperscript{**}   & 1.3559 & -4.0699\textsuperscript{**}  & 1.3559 & -2.6317 & 2.1184  & -2.6313   & 0.4553 \\
$\alpha_{\text{cycling-time}}$      & -6.3078\textsuperscript{***}   & 1.1257 & -6.3078\textsuperscript{***}  & 1.1257 & -3.9240\textsuperscript{**}  & 1.2912  & -4.6405   & 1.0172\\
$\alpha_{\text{walking-time}}$       &-6.8432\textsuperscript{***}   & 0.6762 & -6.8432\textsuperscript{***}  & 0.6762 & -5.6200\textsuperscript{***}  & 1.2597 &-6.2339   & 0.6315\\
$\alpha_\text{traffic}$      & -6.6577\textsuperscript{**}   & 2.2360  & -6.6577\textsuperscript{**}  & 2.2360  & -4.0050\textsuperscript{***}  & 1.1174  & -5.1859   & 1.0467 \\
$C_\text{public}$        & 2.4519\textsuperscript{***}    & 0.5393 & 2.4519\textsuperscript{***}    & 0.5393 & 3.2286\textsuperscript{***}   & 0.9637      & 1.7403   & 1.4935  \\
$C_\text{cycling}$      & 0.6759   & 0.5050  & 0.6759   & 0.5050  & 1.6873\textsuperscript{*}   & 0.8123  & 0.2730   & 0.1501  \\
$C_\text{walking}$       & 4.0373\textsuperscript{***}    & 0.5302 & 4.0373\textsuperscript{***}   & 0.5302 & 4.7248\textsuperscript{***}   & 1.2336  & 3.5505   & 0.4235  \\
\hline
\multicolumn{9}{c}{Modelling error} \\
 & CEL & GMPCA & CEL & GMPCA & CEL & GMPCA & CEL & GMPCA\\
Validation set & 0.5765 & 0.5619 & 0.5765 & 0.5619 & 0.5664 & 0.5676 & 0.5857 & 0.5567\\
Test set & 0.5834 & 0.5580 & 0.5834 & 0.5580 & 0.5916 & 0.5534 & 0.6042 & 0.5465\\
\bottomrule
\end{tabular}
\caption{Modelling estimates (the London model, Level-1)}
\label{tab: Case2level1}
\begin{tablenotes}
      \scriptsize
      \item 
      \begin{flushleft}
      \emph{Note.} \textsuperscript{*} $p <0.05$; \textsuperscript{**} $p <0.01$; \textsuperscript{***} $p <0.001$;   Behavioural inconsistent parameter\slash price ratio: \textcolor{blue}{\textsuperscript{!}}: 2 orders of magnitude deviation from the prior model; \textcolor{blue}{\textsuperscript{!!}}: 3 orders of magnitude deviation.
      \end{flushleft}
    \end{tablenotes}
\end{table}
\begin{figure}[!htbp]
\centering
\includegraphics[%
width=\textwidth]{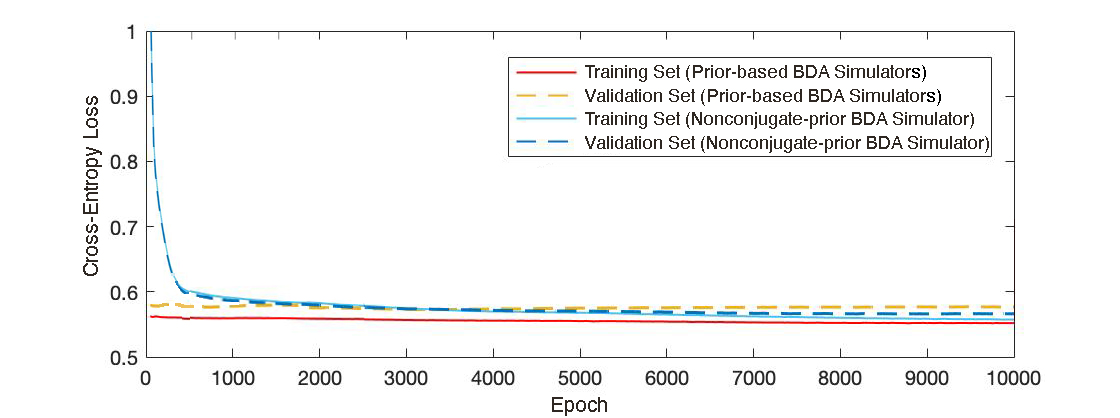}
\caption{\label{fig: Case2level1} Comparison of \acrfull{cel} of the \emph{conjugate-prior}-based \gls{bda} simulators and the \emph{Nonconjugate-prior Bayesian} Simulator (the London model, Level-1)}
\end{figure}

The training of Level-1 London model uses 613 samples.
As shown in Figure \ref{fig: Case2level1}, \glspl{cel} of all the alternative simulators still progress steadily.
They all outperform the direct application approach in terms of predictive error.
\emph{Early stopping} does not occur to \gls{esbda}. 
The \emph{nonconjugate-prior} simulator unsurprisingly converges much more slowly, not reaching an asymptote until around 1000 epochs. 
Most parameters are statistically significant and their signs are consistent with behavioural expectations.
Meanwhile, the \gls{gmpca} values for the output models are fairly evenly matched, with the two \emph{prior}-based \gls{bda} simulators slightly outperforming with the test data.

Despite fine statistics, the monetary ratio of the output~$\alpha_{\text{change-walking-time}}$ from the \emph{nonconjugate-prior} simulator has deviated markedly (12.34 v.s. 1.16) from the corresponding Level-0 London model value.
Several other time-cost ratios also have deviated by up to 80\% of the previous values.
Though the divergence is not as significant as the Level-1 California model's, the output mixed-logit model in this case may result in incorrect behaviour interpretations and mislead policy-decisions.
In contrast, the estimated parameters combinations of other two \emph{prior-based} simulators indicate a stable behavioural representation of the model.

\subsubsection{Level-2}

\begin{table}[!htbp]
\centering
\scriptsize
\begin{tabular}{l|ll|ll|ll|ll}
\toprule
Simulator  & \multicolumn{2}{p{7.5em}|}{\gls{esbda} Simulator (at the 2110\textsuperscript{th} epoch)} & \multicolumn{2}{l|}{\gls{bda} Simulator} & \multicolumn{2}{p{8.5em}|}{\emph{Nonconjugate-prior Bayesian} Simulator} & \multicolumn{2}{p{8.5em}}{Direct Application Approach}   \\
\hline
\multicolumn{9}{c}{Random coefficient}   \\
Latent  & & & & & &  \\
   & Mean              & StDv            & Mean             & StDv           & Mean                & StDv   & Mean                & StDv                  \\
$\mu_\text{price}$                                              & -0.1974  & 0.9121 & -0.2090  & 1.4562     & -0.2539 & 1.5939   & -1.5132  & 0.3949 \\
$\sigma_\text{price}$                                           & 2.6323  & 2.2431 & 8.2019   & 13.0608    & 9.2272  & 16.7234  & 1.8336   & 0.6433 \\
$\mu_\text{operate}$                                             & -2.4376\textsuperscript{***} & 0.3761 & -7.6778  & 4.2876     & -5.1315 & 2.8804    & -5.8014  & 0.7751  \\
$\sigma_\text{operate}$                                            & 3.5423  & 2.8795 & 33.9150  & 82.7603    & 20.4377 & 57.7694 & 3.1118 & 0.9784 \\
$\mu_\text{range}$                                              & -0.9645 & 0.8876 & -8.9406  & 6.4381     & -7.7546 & 5.7963 & -2.4011  & 0.7858  \\
$\sigma_\text{range}$                                           & 2.6502  & 1.7987 & 41.5634  & 100.0234   & 28.3021 & 141.887  & 2.1558  & 0.8384 \\
$\mu_\text{electric}$                                           & -1.5067 & 1.5765 & -0.0101  & 2.1043     & -0.1724 & 1.9302  & -1.5876  & 0.3424 \\
$\sigma_\text{electric}$                                        & 2.6780\textsuperscript{**}   & 0.9251 & 17.3402  & 51.3003    & 12.6582 & 24.0391 & 1.4239   & 0.5577  \\
$\mu_\text{hybrid}$                                             & 1.0523  & 1.5755 & -0.0819  & 2.3232     & -0.2299 & 1.7389  & 0.5282   & 04048 \\
$\sigma_\text{hybrid}$                                          & 4.4210  & 3.9012 & 20.1810   & 72.1741    & 11.0272 & 30.1289  & 1.2361 & 0.5409 \\
\hline
Simulated & & & & & &  \\
$\beta_{\text{price},n}$                                        & -4.1604  & 8.7820 & -24.8885  & 157.8406   & -34.3783 & 245.7799 & -0.5221  & 0.8996 \\
$\beta_{\text{operate},n}$                                      & -0.1070  & 0.2083 & \textcolor{blue}{-131.2370}\textsuperscript{\textcolor{blue}{!}} & 2692.2057  & \textcolor{blue}{-29.4556}\textsuperscript{\textcolor{blue}{!}} & 452.4227  & -0.0138  & 0.0513\\
$\beta_{\text{range},n}$                                        & 1.5674  & 3.1203 & \textcolor{blue}{\textcolor{blue}{730.7354}\textsuperscript{\textcolor{blue}{!}}}\textsuperscript{\textcolor{blue}{!}} & 19697.7659 & 32.1690  & 693.3362 & 0.2797  & 0.6189\\
$\beta_{\text{electric},n}$                                     & -1.0674 & 1.7843 & \textcolor{red}{0.1453}\textsuperscript{\textcolor{red}{e}}   & 4.1886     & -\textcolor{blue}{0.2052}\textsuperscript{\textcolor{blue}{!}} & 3.6437  & -1.5761 & 1.2672 \\
$\beta_{\text{hybrid},n}$                                       & 1.0965  & 2.0724 & \textcolor{blue}{0.0759}\textsuperscript{\textcolor{blue}{!}}   & 4.5287     & -0.2407 & 3.4049  & 0.5383  & 1.1049  \\
\hline
\multicolumn{9}{c}{Fixed coefficient}  \\
$\alpha_\text{high}$                                            & 0.7528\textsuperscript{*}  & 0.3271 & 0.5807   & 0.5610      & 0.6476  & 0.5601  & 0.0958  & 0.1783 \\
$\alpha_\text{midhigh}$                                         & 0.6046  & 0.3906 & \textcolor{blue}{0.0517}\textsuperscript{\textcolor{blue}{!}}   & 0.5007     & \textcolor{blue}{0.0274}\textsuperscript{\textcolor{blue}{!}}  & 0.5095                                    & 0.8014 & 0.1928   \\
\hline
\multicolumn{9}{c}{ Modelling error } \\
  & CEL & GMPCA & CEL & GMPCA & CEL & GMPCA & CEL & GMPCA \\
Validation set & 1.1391 & 0.3201 & 1.1872 & 0.3050 & 1.1903 &0.3041 & 1.1992 & 0.3014\\
Test set & 1.4465 & 0.2354 & 1.5120 & 0.2205 & 1.9494 & 0.1424 & 1.9754 & 0.1387 \\
\bottomrule
\end{tabular}
\caption{Modelling estimates (the California Model, Level-2)}
\label{tab: Case1level2}
\begin{tablenotes}
      \scriptsize
      \item 
      \begin{flushleft}
      \emph{Note.} \textsuperscript{*} $p <0.05$; \textsuperscript{**} $p <0.01$; \textsuperscript{***} $p <0.001$; 
\newline Behavioural inconsistent parameter/price ratio:  \textcolor{blue}{\textsuperscript{!}}: 2 orders of magnitude deviation from the prior model; \textcolor{blue}{\textsuperscript{!!}}: 3 orders of magnitude deviation.
      \end{flushleft}
    \end{tablenotes}
\end{table}
\begin{figure}[!htbp]
\centering
\includegraphics[%
width=\textwidth]{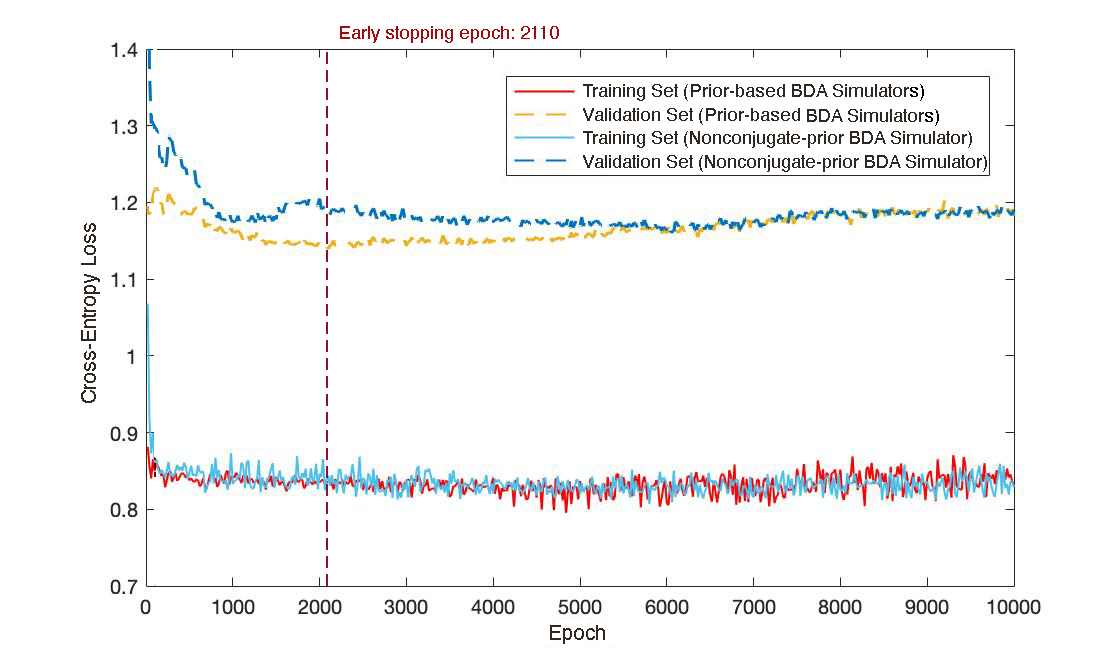}
\caption{\label{fig: Case1level2} Comparison of \acrfull{cel} of the \emph{conjugate-prior}-based \gls{bda} simulators and the \emph{Nonconjugate-prior Bayesian} Simulator (the California model, Level-2). \emph{Note:} there is a big gap between training and validation sets in performance, even at epoch 0 for the prior-based \gls{bda} - this is a result of sampling noise in the very small test and validation samples.}
\end{figure}

\begin{table}[!htbp]
\centering
\scriptsize 
\begin{tabular}{l|ll|ll|ll|ll}
\toprule
Simulator & \multicolumn{2}{p{7.5em}|}{\gls{esbda} Simulator (at the 80\textsuperscript{th} epoch)} & \multicolumn{2}{p{6.5em}|}{\gls{bda} Simulator} & \multicolumn{2}{p{8.5em}|}{\emph{Nonconjugate-prior Bayesian} Simulator}  & \multicolumn{2}{p{8.5em}}{Direct Application Approach}    \\
\hline
\multicolumn{9}{c}{Random coefficient}   \\
Latent  & & & & & &  \\
   & Mean              & StDv            & Mean             & StDv           & Mean                & StDv     & Mean                & StDv             \\
$\mu_{\text{cost}}$                                        & -0.1613\textsuperscript{***}  & 0.0030 & -0.1515\textsuperscript{***}  & 0.0200   & -0.1567\textsuperscript{***} & 0.0116 & -0.1531   & 0.0409\\
$\sigma_{\text{cost}}$                                     & 0.0040\textsuperscript{***}   & 0.0002 & 0.0241   & 0.0126 & 0.01930\textsuperscript{*}  & 0.0077 & 0.0149   & 0.0275 \\
Simulated & & & & & &  \\
$\beta_{\text{cost},n}$                                      & -0.1670 & 0.0627 & -0.1656  & 0.1535 & -0.1693 & 0.1376    & -0.1641  & 0.2182     \\
\hline
\multicolumn{9}{c}{Fixed coefficient} \\
$\alpha_{\text{driving-time}}$                                  & -2.9900\textsuperscript{***}  & 0.1950 & -1.0078  & 2.3048\textsuperscript{*} & -5.2698 & 2.1117  &-4.2537   & 1.5895\\
$\alpha_{\text{access-time}}$                                   & -3.8152\textsuperscript{***}  & 0.3183 & -3.9098\textsuperscript{*}  & 1.5454 & -2.4823 & 2.3405 & -5.7911   & 1.2652 \\
$\alpha_{\text{bus-time}}$                                      & -2.5568\textsuperscript{***} & 0.6217 & -1.2926  & 0.9609 & -1.8107 & 1.7206 & -3.4614   & 1.0902  \\
$\alpha_{\text{rail-time}}$                                     & -2.3364\textsuperscript{***} & 0.1017 & -0.3130  & 1.8523 & -0.8199 & 2.9039   & -4.2861   & 1.6784\\
$\alpha_{\text{change-walking-time}}$                           & -2.3003\textsuperscript{***} & 0.1561  & -0.7313  & 1.6394 & -0.0641 & 1.0505   & -4.8911    & 2.4391\\
$\alpha_{\text{change-waiting-time}}$                           & -2.5137\textsuperscript{***} & 0.1764 & -3.3472  & 1.8070 & -2.6195 & 2.6344 & -4.0699   & 1.3559\\
$\alpha_{\text{cycling-time}}$                                  & -5.0906\textsuperscript{***}  & 0.2034 & -4.5897\textsuperscript{***}  & 1.0737 & -5.0841 & 2.8613 & -6.3078  & 1.1257 \\
$\alpha_{\text{walking-time}}$                                  & -6.7379\textsuperscript{***}  & 0.2872 & -8.5262\textsuperscript{***}  & 1.2530  & -7.5704\textsuperscript{*} & 2.1896  &-6.8432   & 0.6762 \\
$\alpha_\text{traffic}$                                                  & -5.3766\textsuperscript{**} & 0.3347  & -10.1548\textsuperscript{***} & 2.8023 & -9.4864\textsuperscript{**} & 3.4395 & -6.6577   & 2.2360 \\
$C_\text{public}$                                                  & 1.7731\textsuperscript{***}   & 0.1849 & 1.8745\textsuperscript{*}   & 0.7839 & 1.6103  & 0.9444 & 2.4519    & 0.5393\\
$C_\text{cycling}$                                                  & 0.4090\textsuperscript{***}    & 0.1131 & 0.8278   & 0.6686 & 0.8456  & 0.8890  & 0.6759   & 0.5050 \\
$C_\text{walking}$                                                  & 3.6822\textsuperscript{***}   & 0.2830  & 5.0738\textsuperscript{***}   & 0.7619 & 4.4023\textsuperscript{***}  & 1.2089  & 4.0373     & 0.5302  \\
\hline
\multicolumn{9}{c}{Modelling error } \\
 & CEL & GMPCA & CEL & GMPCA & CEL & GMPCA & CEL & GMPCA\\
Validation set & 0.5368 & 0.5845 & 0.5517 & 0.5760 & 0.5489 & 0.5776 & 0.5493 & 0.5773\\
Test set & 0.6164 & 0.5398 & 0.8928 & 0.4095 & 1.1476 & 0.3174 & 0.8732 & 0.4176\\
\bottomrule
\end{tabular}
\caption{Modelling estimates (the London model, Level-2)}
\label{tab: Case2level2}
\begin{tablenotes}
      \scriptsize
      \item 
      \begin{flushleft}
      \emph{Note.} \textsuperscript{*} $p <0.05$; \textsuperscript{**} $p <0.01$; \textsuperscript{***} $p <0.001$; \textsuperscript{\textcolor{red}{e}} \textcolor{red}{sign-error};  Behavioural inconsistent parameter\slash price ratio:  \textcolor{blue}{\textsuperscript{!}}: 2 orders of magnitude deviation from the prior model; \textcolor{blue}{\textsuperscript{!!}}: 3 orders of magnitude deviation.
      \end{flushleft}
\end{tablenotes}
\end{table}
\begin{figure}[!htbp]
\centering
\includegraphics[%
width=\textwidth]{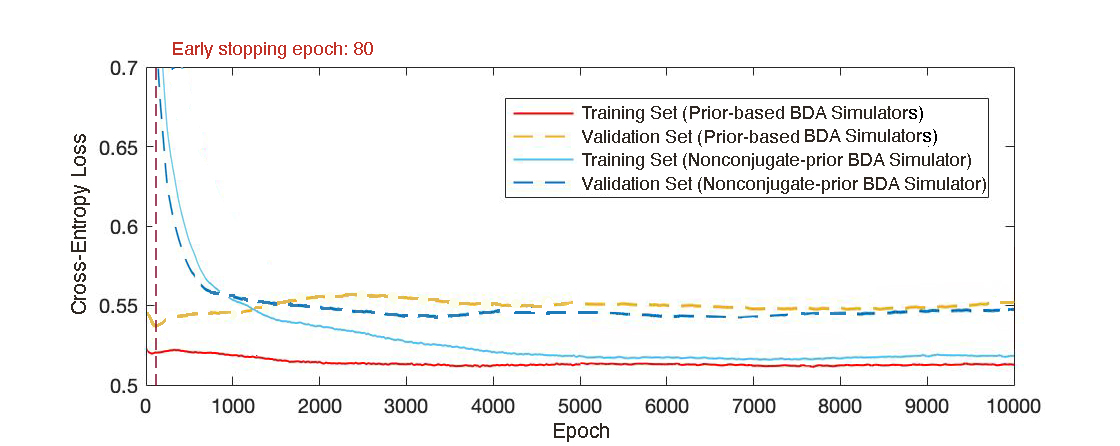}
\caption{\label{fig: Case2level2}Comparison of \acrfull{cel} of the \emph{conjugate-prior}-based \gls{bda} simulators and the \emph{Nonconjugate-prior Bayesian} Simulator (the London model, Level-2)}
\end{figure}

The benchmark simulators both show increased \gls{cel} fluctuations in modelling at Level 2 (Figure \ref{fig: Case1level2} \& Figure \ref{fig: Case2level2}).
With a handful of 75 choice training samples, the California model undergoes intense volatility, and \gls{esbda} therefore encounters \emph{early stopping} at the 2110\textsuperscript{th} epoch.

Difference between the simulators' modelling errors becomes marginal as the simulation continues, as shown by Figure \ref{fig: Case1level2}.
In the validation set and the out-of-sample test of the California model, however, the advantage of \gls{esbda}'s output model is clear in \gls{gmpca}.
For all simulators, there is a large difference between the training and validation/test performances (as indicated by the \gls{cel} and \gls{gmpca} scores). 
This is due to the high relative sampling noise in the small sample sizes for the validation and test sets. 
For the \gls{bda} and \emph{Nonconjugate prior Bayesian} simulators, this sampling noise results in model overfitting during estimation, with resulting  parameter values inconsistent with behavioural theory.
The  monetary ratios of some parameters' vary by over 3 orders of magnitude from the Level-1 model, \gls{bda}'s parameter~$\beta_{n\_\text{electric}}$, additionally has an incorrect sign. 
This is despite the model being applied to the same context as the Level-1 model and the training data randomly drawn from the Level-1 data. 

This shows a clear limit of both the nonconjugate-prior and \gls{bda} simulators in terms of sample size required. 
Hence the corresponding benchmark models are considered as invalid behavioural models.
For the \gls{esbda} simulator, \emph{early stopping} prevents the model from overfitting during model training, 
and the final model parameters are consistent with the Level-1 estimates and behavioural expectations. 

Unlike the California model, given a total of 266 training samples, the Level-2 London model has much steadier modelling progress and much less time-cost ratio deviation.
Yet the training and validation model's prediction performances do not differ much, \gls{esbda} shows a clear advantage in its out-of-sample prediction, \glspl{gmpca} being 0.5398 v.s. 0.4095 v.s. 0.3174 v.s. 0.4176.

\subsubsection{Level-3}
The sample size for the Level-2 California modelling is already very small (merely 5 individuals' data for training), which does not support subdivision for Level-3 experimentation.
As such, Level-3 modelling is carried out for the London case-study only.

As with the Level-2 scenario for the California case-study, the sampling noise from the very small sample size for the Level-3 London model results in large discrepancies in train and test performance for the nonconjugate-prior estimator as well as the \gls{bda} without \emph{early stopping}. 
This indicates that overfitting occurs during model estimation. Once again, the \emph{early stopping} procedure in the \gls{esbda} simulator prevents this overfitting, 
and results in a final model with substantially higher test performance. 
Moreover, the overfit \gls{bda} and noncojugate-prior models have substantial sign and scale errors in the parameters resulting from the overfitting, whilst the \gls{esbda} maintains full consistency with behavioural expectations.

\begin{table}[!htbp]
\centering
\scriptsize 
\begin{tabular}{l|ll|ll|ll|ll}
\toprule
Simulator & \multicolumn{2}{p{7.5em}|}{\gls{esbda} Simulator (at the 60\textsuperscript{th} epoch)} & \multicolumn{2}{p{6.5em}|}{\gls{bda} Simulator} & \multicolumn{2}{p{8.5em}|}{\emph{Nonconjugate-prior Bayesian} Simulator}  & \multicolumn{2}{p{8.5em}}{Direct Application Approach}    \\
\hline
\multicolumn{9}{c}{Random coefficient}   \\\
Latent  & & & & & &  \\
   & Mean              & StDv            & Mean             & StDv           & Mean                & StDv        & Mean                & StDv           \\
$\mu_{\text{cost}}$                                        & -0.2219\textsuperscript{**} & 0.0706 & -0.3780   & 0.6723 & -0.6685 & 0.6493 & -0.1613  & 0.0030\\
$\sigma_{\text{cost}}$                                     & 0.0770\textsuperscript{***}   & 0.0056 & 0.8515   & 1.0724 & 1.1269  & 1.5951  & 0.0040   & 0.0002 \\
Simulated & & & & & &  \\
$\beta_{\text{cost},n}$                                      & -0.2470 & 0.2746 & -0.4616  & 0.9136 & -0.7647 & 1.0510   & -0.1670 & 0.0627 \\
\hline
\multicolumn{9}{c}{Fixed coefficient} \\
$\alpha_{\text{driving-time}}$                                  & -3.4202\textsuperscript{***} & 0.1423 & \textcolor{red}{2.1181}\textsuperscript{\textcolor{red}{e}}   & 2.7649 & -5.7653 & 2.7036 & -2.9900   & 0.1950  \\
$\alpha_{\text{access-time}}$                                   & -3.3099\textsuperscript{***} & 0.2615 & \textcolor{red}{3.6504}\textsuperscript{\textcolor{red}{e}}   & 3.1918 & \textcolor{red}{3.9425}\textsuperscript{\textcolor{red}{e}}  & 2.1141 & -3.8152  & 0.3183 \\
$\alpha_{\text{bus-time}}$                                      & -2.5795\textsuperscript{***} & 0.1886 & \textcolor{red}{3.0410}\textsuperscript{\textcolor{red}{e}}    & 3.0798 & \textcolor{red}{7.1499}\textsuperscript{\textcolor{red}{e}}  & 3.8681 & -2.5568 & 0.6217\\
$\alpha_{\text{rail-time}}$                                     & -2.2033\textsuperscript{***} & 0.2303 & -1.2485  & 3.1867 & \textcolor{red}{12.5610}\textsuperscript{\textcolor{red}{e}}  & 6.4642 & -2.3364  & 0.1017\\
$\alpha_{\text{change-walking-time}}$                           & -0.8872 & 0.6649 & -1.8847  & 4.8146 & -1.2363 & 2.4461  & -2.3003 & 0.1561\\
$\alpha_{\text{change-waiting-time}}$                           & -2.4913\textsuperscript{***} & 0.3341 & \textcolor{red}{5.0345}\textsuperscript{\textcolor{red}{e}}  & 5.8134 & \textcolor{red}{5.5778\textsuperscript{*}}\textsuperscript{\textcolor{red}{e}}  & 2.2320  & -2.5137  & 0.1764\\
$\alpha_{\text{cycling-time}}$                                  & -5.2001\textsuperscript{***} & 0.5262 & -5.2188\textsuperscript{*}  & 1.7910  & \textcolor{red}{2.7879\textsuperscript{*}}\textsuperscript{\textcolor{red}{e}}  & 2.4994 & -5.0906   & 0.2034 \\
$\alpha_{\text{walking-time}}$                                  & -6.2652\textsuperscript{***} & 0.2117 & -13.9090\textsuperscript{*}  & 4.7416 & -5.5405 & 3.1127 & -6.7379  & 0.2872\\
$\alpha_\text{traffic}$                                                  & -5.2321\textsuperscript{***} & 0.3069 & -14.5173\textsuperscript{*} & 4.7514 & -4.7229\textsuperscript{*} & 1.7556 & -5.3766  & 0.3347\\
$C_\text{public}$                                                  & 1.6822\textsuperscript{***}  & 0.2160  & 4.8957\textsuperscript{**}   & 1.4114 & \textcolor{red}{-0.4629}\textsuperscript{\textcolor{red}{e}} & 2.1898  & 1.7731   & 0.1849\\
$C_\text{cycling}$                                                  & 0.8333  & 0.4028 & 5.3572\textsuperscript{*}   & 1.9971 & -1.4617 & 1.8743 & 0.4090   & 0.1131\\
$C_\text{walking}$                                                  & 3.5707\textsuperscript{***}  & 0.4112 & 11.0504\textsuperscript{*}  & 3.8048 & 1.9144  & 1.1669  & 3.6822     & 0.2830\\
\hline
\multicolumn{9}{c}{Modelling error } \\
 & CEL & GMPCA & CEL & GMPCA & CEL & GMPCA & CEL & GMPCA\\
Validation set  & 0.5869 & 0.5561  & 1.1364 & 0.3209 & 0.9305 & 0.3944 & 0.6089 & 0.5439\\
Test  set & 0.6635 & 0.5150 & 1.2883 & 0.2757 & 1.0905 & 0.3360 & 0.6971 & 0.4980\\
\bottomrule
\end{tabular}
\caption{Modelling estimates (the London model, Level-3)}
\label{tab: Case2level3}
\begin{tablenotes}
      \scriptsize
      \item 
      \begin{flushleft}
      \emph{Note.} \textsuperscript{*} $p <0.05$; \textsuperscript{**} $p <0.01$; \textsuperscript{***} $p <0.001$; \textsuperscript{\textcolor{red}{e}} \textcolor{red}{sign-error};    Behavioural inconsistent parameter\slash price ratio: \textcolor{blue}{\textsuperscript{!}}: 2 orders of magnitude deviation from the prior model; \textcolor{blue}{\textsuperscript{!!}}: 3 orders of magnitude deviation.
      \end{flushleft}
\end{tablenotes}
\end{table}

Overall, \gls{esbda} shows clear advantage over the benchmark simulators in quality of estimates, steadiness of the modelling process, the generally higher \gls{gmpca} and a major merit in deriving behavioural interpretative and insightful mixed-logit models.
These advantages are gradually manifested as sample size reduces.

Modelling results clearly demonstrate the value of the early-stopping procedure when dealing with very small sample sizes. 
Note that each modelling level represents a certain demographic group in the London study-case. 
As such, the estimated models represent true transferred models from the corresponding prior models.
Robust performance of the \gls{esbda} in the London study-case therefore demonstrates the ability of the \gls{bda} and \gls{esbda} to achieve better model transfers, 
and the ability of the \gls{esbda} to prevent overfitting with very small training samples.
Therefore, these results illustrate the potential of \gls{esbda} in model transfer.

\glsresetall
\section{Conclusions and Further Work}
\label{sec: con}
This study advances the basic \emph{Bayesian} simulator of Train \& Garrett \citetext{2006} to enable new contexts with lower data availability to establish their models by transferring from an existing mixed-logit model of another context with rich data.
Through the combination of \emph{Bayesian} modelling and \gls{ml} techniques, the modelling of the proposed \gls{esbda} simulator incorporates a previously established parameters combination as an informative \emph{conjugate-prior} and assimilates collected data through iterative \emph{Bayesian inference}.
Data assimilation helps the resultant model avoid the under-fitting problem caused by naive application of any empirical model that is not tailored to the new context.
The \emph{early stopping} procedure, on the other hand, prevents the modelling from over-fitting or non-convergence, which are two recurrent problems in small sample modelling.
Meanwhile, through the \emph{early stopping} procedure, a lightweight \gls{msl} analogue is equipped to complement the \emph{Bayes procedure}.
\gls{esbda} has therefore consolidated the merits of these two most prominent mixed-logit simulators.

\gls{esbda} is benchmarked against the direct application approach and two reference simulators --- a \emph{non-conjugate} simulator and a \gls{bda} which has no \emph{early stopping} trigger.
The modelling study consists of experiments of two mixed-logit models and at multiple levels of sample size.
Comparisons are made on (i) estimation statistics, i.e., statistical significance, in-sample and out-of-sample prediction errors; (ii) behavioural consistency of the estimated mixed-logit models; and (iii) steadiness of modelling process.

The output model of \gls{esbda} outperforms its counterparts of the benchmark simulators in each of the three above dimensions in every experiment.
The results also indicate the high behavioural consistency and strong explanatory power of the output models from \gls{esbda} compared with the benchmark operators. 
Another advantage of \emph{prior-based Bayesian} approaches (including \gls{bda} and \gls{esbda}) is that they can avoid unreasonable variation in model estimates arising from random initial parameter estimates. 
Overall, the results in this paper indicate the \gls{esbda} simulator could be used as a practical, economical and relatively time-saving tool to assist in analysing choice behaviour, particularly for modelling specific population groups and for future estimation with lower data availability.

The \gls{esbda} simulator in its current state has several limitations that direct future studies to fulfilling its full potential.
Firstly, we would like to explore technical updates which may reinforce the simulator, e.g., the \emph{Cross-Validation} and \emph{Hamiltonian Monte Carlo} techniques.
Another direction for future study is to expand the adaptation of the \gls{esbda} simulator to e.g., (i) mixed-logit models with a flexible mixing distribution, (ii) other types of classic \glspl{dcm} and (iii) extended \glspl{dcm} \citep{walker2001extended}, such as integrated framework, flexible error structures, and latent variables.

An important future work is to expand the applications of the \gls{esbda} simulator.
It is of our particular interest to develop (i) transferred models for sub-populations from the full population model, which helps to investigate heterogeneous choice behaviour between demographic groups and (ii) a \emph{Bayesian} framework that continuously updates the future travel behaviour estimation model as new data becomes available, which accommodates the need to investigate post-pandemic travel behaviour.

\section{Acknowledgements} 

This research was conducted as part of Shanshan Xie's doctoral studies, which are funded by the China Scholarship Council.

We are grateful to Kenneth Train, as the \emph{Nonconjugate prior Bayesian} Simulator, which is our principle benchmark simulator, is adapted from his \gls{hb} procedure codes \citep{Train2006}.
In addition, we sincerely thank Kenneth Train and Garrett Sonnier for providing the dataset of the California case-study.
We would also like to thank Jamil Nur for his data processing work for the London travel mode choice data.

\bibliographystyle{elsarticle-harv}
\bibliography{BDAModel.bib}

\end{document}